\documentclass[acmtog]{acmart}
\acmSubmissionID{198}

\usepackage{booktabs} 

\usepackage[ruled]{algorithm2e} 

\SetAlFnt{\small}
\SetAlCapFnt{\small}
\SetAlCapNameFnt{\small}
\SetAlCapHSkip{0pt}
\usepackage{multirow}
\usepackage{bm}

\setcopyright{acmcopyright}\acmJournal{TOG}
\acmYear{2020}\acmVolume{39}\acmNumber{6}\acmArticle{232}\acmMonth{12} \acmDOI{10.1145/3414685.3417796}

\usepackage{color}
\definecolor{turquoise}{rgb}{0.6,0.4,0}
\definecolor{purple}{rgb}{0.65,0,0.65}
\definecolor{dark_green}{rgb}{0, 0.5, 0}
\definecolor{green}{rgb}{0, 0.8, 0}
\definecolor{orange}{rgb}{0.8, 0.6, 0.2}
\definecolor{red}{rgb}{0.9, 0, 0}
\definecolor{brown}{rgb}{0.5, 0.16, 0.16}

\newcommand{\rh}[1]{{\color{black}#1}}

\newcommand{\change}[1]{{\color{black}#1}}

\newcommand{\tapnet}{{\sc TAP-Net}}

\begin{document}

\title{TAP-Net: Transport-and-Pack using Reinforcement Learning}

\author{Ruizhen Hu}
\affiliation{%
	\department{College of Computer Science \& Software Engineering}
	\institution{Shenzhen University}
}
\email{ruizhen.hu@gmail.com}

\author{Juzhan Xu}
\affiliation{%
	\institution{Shenzhen University}
}

\author{Bin Chen}
\affiliation{%
	\institution{Shenzhen University}
}

\author{Minglun Gong}
\affiliation{%
	\institution{University of Guelph}
}

\author{Hao Zhang}
\affiliation{%
	\institution{Simon Fraser University}
}
\author{Hui Huang}
\authornote{Corresponding author: Hui Huang (hhzhiyan@gmail.com)}
\affiliation{%
	\department{College of Computer Science \& Software Engineering}
	\institution{Shenzhen University}
}

\renewcommand\shortauthors{R. Hu, J. Xu, B. Chen, M. Gong, H. Zhang, and H. Huang}

\begin{teaserfigure}
	\includegraphics[width=0.98\textwidth]{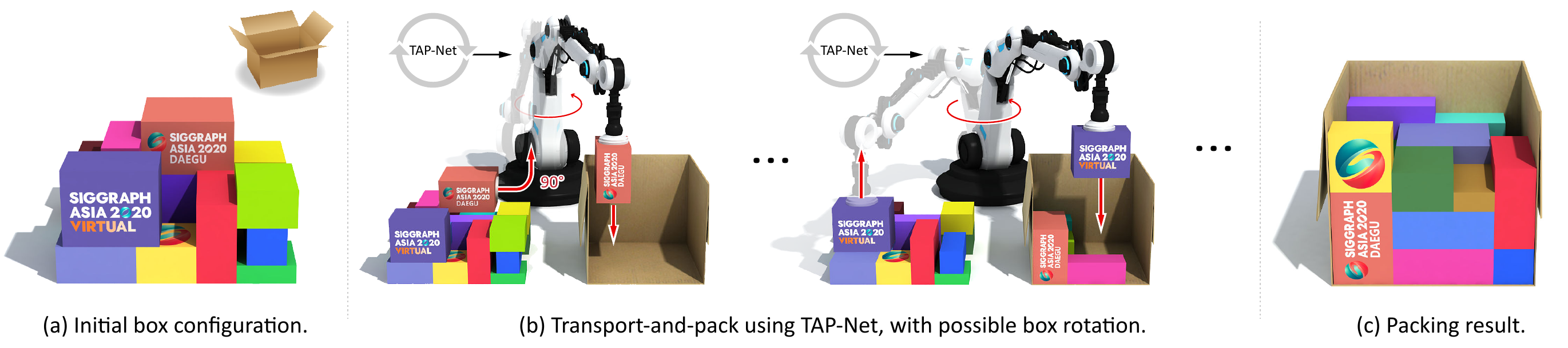}
	 \centering
	\caption{Given an initial spatial configuration of boxes (a), our neural network, \tapnet{}, iteratively transports and packs (b) the boxes compactly into a target container (c). \tapnet{} is trained to output a packing order and box orientations ($90^\circ$ rotations are allowed; see the orange box with the SIGGRAPH Logo) through reinforcement learning, under obstruction and accessibility constraints, and with rewards designated to facilitate compact packing.}
	\label{fig:teaser}
\end{teaserfigure}

\begin{abstract}
We introduce the {\em transport-and-pack\/} (TAP) problem, a frequently encountered instance of 
real-world packing, and develop a {\em neural optimization\/} solution based on {\em reinforcement learning\/}.
Given an initial spatial configuration of boxes, we seek an efficient method to iteratively transport
and pack the boxes compactly into a target container. Due to obstruction and accessibility constraints, our 
problem \rh{has to add a new search dimension, i.e., finding an optimal {\em transport sequence\/},} to the 
already immense search space for packing alone. Using a learning-based approach, a trained network can 
learn and encode solution patterns to guide the solution of new problem instances instead of executing an 
expensive online search. In our work, we represent the transport constraints using a {\em precedence graph\/} 
and train a neural network, coined \tapnet{}, using reinforcement learning to reward {\em efficient\/} and 
{\em stable\/} packing. The network is built on an encoder-decoder architecture, where the encoder employs 
convolution layers to encode the box geometry and precedence graph and the decoder is a recurrent neural 
network (RNN) which inputs the current encoder output, as well as the current box packing state of the target 
container, and outputs the next box to pack, as well as its orientation. 
We train our network on {\em randomly generated\/} initial box configurations, 
{\em without supervision\/}, via policy gradients to learn optimal TAP policies to maximize packing efficiency and stability.
We demonstrate the performance of \tapnet{} on a variety of examples, evaluating the network through ablation
studies and comparisons to baselines and \rh{alternative network designs}. We also show that our network generalizes
well to larger problem instances, when trained on small-sized inputs.
%

\end{abstract}

%
%


%
%


\maketitle

\section{Introduction}
\label{sec:intro}

Packing is a well-known discrete optimization problem that has found compelling geometry
applications in computer graphics, e.g., texture atlas generation~\cite{carr2002,noll11,limper18_boxcutter}, 
artistic layout~\cite{reinert2013}, 2D panel fabrication~\cite{saakes13,limper18_boxcutter}, \change{the Escherization problem~\cite{nagata2020}, jigsaw puzzles~\cite{wei2019}, mosaic stylization~\cite{doyle2019}} and 3D printing~\cite{chen15_dapper}.
While these applications only need to optimize the {\em packing efficiency\/} of {\em virtual\/} object arrangements for
display, storage, and fabrication, \rh{real-world applications, such as robot-assisted
packaging and transport, often must face additional constraints arising from the {\em physical process\/} of packing.}

\begin{figure*}[!t]
    \centering
    \includegraphics[width=0.99\textwidth]{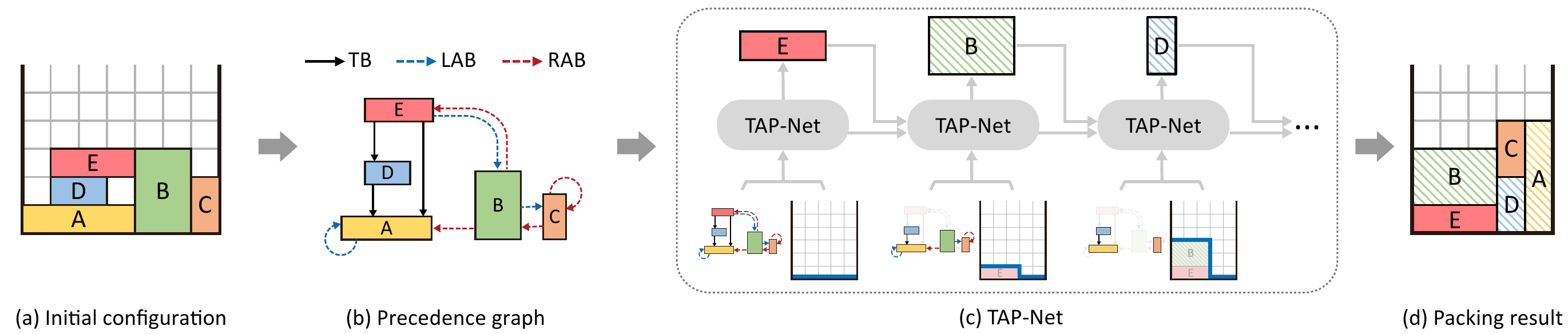}
\caption{
Overview of our TAP framework, illustrated using a 2D example. Given an initial layout of boxes (a), we first analyze the transport constraints (i.e., obstructions and accessibility) which would dictate packing orders. For example, TB means ``Top Blocking'' of box D by box E, so D can only be packed after E; see more details in Figure~\ref{fig:access} and Section~\ref{sec:overview}. We represent these constraints in a precedence graph (b). The precedence graph and the initial state of the target container are fed into the network \tapnet{} to determine which box should be packed first (c). Updated precedence graph and packing state of the target container are iteratively fed to the network to determine subsequent packing order until all boxes are packed (d).}
\label{fig:overview}
\end{figure*}

\rh{In a frequently encountered instance of real-world packing, the objects to be packed are already in a physical arrangement,}
e.g., a packed storage state as inventory has been accumulated over time. In this case, the object movement must be
\rh{{\em sequential} and respect a} {\em partial order\/}, e.g., an object cannot be packed unless all objects on top of it 
have been cleared. Hence, while packing alone is already a difficult combinatorial optimization problem, the new 
problem instance adds an extra search dimension to \rh{find an optimal {\em transport sequence\/}}, making it a 
{\em transport-and-pack\/} (TAP) problem.

To address the compounded computational challenges of the TAP problem, we propose a {\em neural combinatorial
optimization\/} approach using {\em reinforcement learning\/} (RL). Solving hard optimizations using
neural networks exhibits a new trend in machine learning. Such approaches exploit learnable patterns 
in the problems to enable trained networks to solve new problem instances with greater efficiency than existing 
approximation algorithms and better quality and generality than heuristic search. \rh{Our motivations for adopting 
RL are two-fold. First, RL is a natural fit for TAP since the problem involves sequential decision making~\cite{Keneshloo2019}. 
Moreover, with RL, the learning is unsupervised, thus avoiding the need for ground-truth TAP solutions that are difficult to
obtain.}

In a first attempt, we transport-and-pack objects abstracted by axis-aligned boxes (AABBs), possibly rotated by $90^\circ$, 
into a target container, which is also a box itself (with unbounded height); see Figures~\ref{fig:overview}(a) and (d). Our goal 
is \rh{to train a neural network, coined \tapnet{}, to infer a TAP sequence so as to} maximize packing 
{\em efficiency\/} and {\em stability\/} while respecting transport constraints. These constraints are modeled using a {\em 
precedence graph\/}, which accounts for precedence relations due to blocking or obstruction from the top
or the sides;
see Figure~\ref{fig:overview} and~\ref{fig:access} for more details. 

\begin{figure}[!t]
	\centering
	\includegraphics[width=0.5\textwidth]{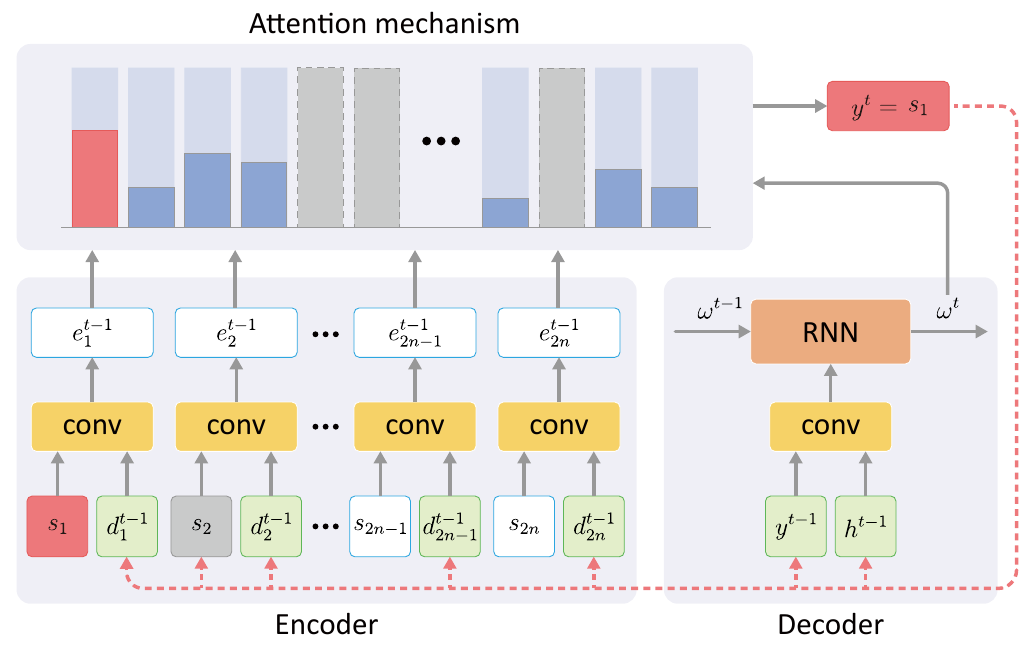}
	\caption{The architecture of \tapnet. Two modules, an encoder and a decoder, are trained to collectively compute a probability map over all valid states under each step using an attention mechanism. The state with maximum probability ($s_1$ in this case) is selected as output for the current step, and used to upated all dynamic information and item mask.}
\label{fig:network}
\end{figure}

\tapnet{} takes as input \rh{a set of AABBs to be packed, each represented by box geometry and a precedence subgraph involving 
the box, as well as the current box packing state of the target container,} and outputs the next box to pack and its orientation 
(i.e., rotated or not). \rh{When trained, \tapnet{} is applied repeatedly to produce a TAP sequence from the initial input box 
arrangement, as shown in Figure~\ref{fig:overview}(c).} The network is built on an encoder-decoder architecture with an attention 
mechanism, as shown in Figure~\ref{fig:network}. Specifically, the encoder employs convolution layers to encode the box 
geometry and precedence information, and the decoder is a recurrent neural network (RNN) which makes inference from the 
current encoder output and the packing state in the target container.

\rh{In the context of RL, \tapnet{} serves as the agent, the sequence of box and orientation selections are the actions, and 
the state is given by the {\em joint status\/} of the transport (i.e., precedence graph) and packing (i.e., box
arrangements in the target container) components of our problem.} The rewards are defined based on the efficiency and 
stability with which the selected oriented boxes are packed into the container. 
\tapnet{} is trained on {\em randomly generated\/} initial box configurations, {\em without supervision\/}, via policy 
gradients to learn optimal TAP policies to maximize packing efficiency and stability.
%
We demonstrate the performance of \tapnet{} on a variety of 2D and 3D examples, evaluating the network through ablation studies 
and comparisons to \rh{baselines and alternative network designs.}


\rh{
Compared to state-of-the-art neural networks and RL-based approaches to solve hard optimization problems, our problem and network
design offer several novel features:}
\begin{itemize}
\item 
\rh{First, our RL states are not pre-determined: both the precedence graph and the packing
state are {\em dynamic\/} --- they change over time as boxes are transported and packed.} This problem setting precludes 
direct applications of well-known neural optimization models such as Pointer Networks~\cite{vinyals15,bello17}, which have 
been employed to tackle hard problems including the Traveling Salesman (TSP).
\item
Second, \rh{unlike typical RL- or RNN-based approaches to sequence generation
\cite{Keneshloo2019,nazari18}, where the network is only tasked to produce a sequence,
TAP must integrate two tasks: box selection, which results in a sequence, {\em and\/} packing. 
In addition, box selection is directly impacted by how the boxes are packed.}
Hence, in our network, we perform packing after every box selection and feed the updated packing state back to the network 
to select the next box. As shown in Section~\ref{sec:results}, \rh{training \tapnet{} with such {\em intermediate\/} packing states
significantly improves the results, when compared to the alternative of packing the boxes only {\em after\/} the entire sequence has been
generated.}
\item
At last, the {\em incrementality\/} of our solution framework allows \tapnet{} to generalize well to larger problem instances 
when trained on smaller-sized inputs, \rh{while conventional wisdom on RL and deep RL frameworks often points to their 
limited generalization capabilities~\cite{packer2018,cobbe2018}.}
In our work, as \tapnet{} incrementally builds the action sequence, at each step, we can judiciously limit the size of the input to the network
so that it is inline with that from the training stage. We can apply a ``rolling'' scheme to progressively update the 
precedence graph: as an object is packed, it is removed from the graph and one new object comes in; see Section~\ref{sec:extend_larger} 
for details.
\end{itemize}

\rh{Note that by design, \tapnet{} is {\em not\/} trained via RL to pack the selected boxes. Instead, we employ simple heuristic-based 
strategies to obtain packing to evaluate the reward. Our intuition is that it is difficult for the network to learn effective box selection
{\em and\/} packing policies together based only on our simple packing reward. This is verified by our experiments which show that 
employing heuristic packing policies consistently outperforms RL frameworks, trained end-to-end, which incorporate a packing network.}

\section{Related work}
\label{sec:related}

In this section, we first discuss variations of the packing problem that have been studied in computer graphics.
We then go over traditional approaches to solve these problems. Most closely related to our work are recent
techniques developed in the field of machine learning, on neural optimization models. We cover the most relevant works and reveal how they have inspired our approach.

\paragraph{Packing in graphics.}
To store and access surface signals such as colors, textures, or normals efficiently, the chartification problem aims to maximize the packing efficiency of a set of parameterized 2D charts~\cite{levy2002least,noll11,sander2003multi}. To minimize material waste during fabircation, Ma et al.~\shortcite{ma2018packing} propose a method to place as many given irregular objects as possible in a non-overlapping configuration within a given container. \change{The Escherzation problem~\cite{kaplan2000,nagata2020}, on the other hand, tries to find one single closed figure that is as close as possible to the input figure that tile the plane. Another related and interesting problem is jigsaw puzzles, the goal of which is to recover the correct configuration of a set of regular but disordered image patches. }

Aside from those works focusing purely on packing, a recent emerging problem is {\em decompose-and-pack\/}, where the shapes to be packed are decomposed or cut from an input object. Hence, shape decomposition and packing must be jointly optimized to maximize packing efficiency. For chart packing, Limper et al. ~\shortcite{limper18_boxcutter} allow new cuts to be added on the given charts to obtain the desired trade-off between packing efficiency and boundary length of charts. Liu et al.~\shortcite{liu2019atlas} further improve packing efficiency by converting the original problem to a rectangle packing problem to solve in a more efficient and accurate way. \change{Doyle et al.~\shortcite{doyle2019} presents a method to synthesize pebble mosaic, which decomposes the image into small smooth shapes resembling pebble cross-sections.} For fabrication, Koo et al.~\shortcite{koo2016towards} allow the geometry of 2D furniture pieces to be changed slightly to guide the modification of furniture design with minimal material waste, while Vanek et al.~\shortcite{vanek2014packmerger} and Chen et al.~\shortcite{chen15_dapper} directly optimize the decomposition of the given 3D shape over the surface shell or solid volume so that the decomposed parts can be tightly packed for 3D printing to minimize both printing time and material waste.

Our TAP problem can be seen as a variant of the traditional bin packing problem and it shares the same goal as the other variants in maximizing packing efficiency. The key difference is that all previous works are concerned only with the final packing state instead of the packing {\em process\/}, which is a key to take into account for real-world packing applications, such as transport planning.
\vspace{-3pt}

\paragraph{Packing solutions.}
The traditional bin packing problem is an NP-hard combinatorial problem \cite{chazelle1989complexity}, for which various heuristics have been proposed.
One common strategy is to solve the problem in two steps: determine the packing order and then pack the objects based on some fixed packing strategy.
The packing order can be optimized using genetic algorithms~\cite{ramos16}, simulated annealing~\cite{liu15}, or more advanced RL-based methods~\cite{hu17, duan19}. There are also different choices for the packing strategy, e.g., Deepest-Bottom-Left with Fill (DBLF) strategy~\cite{karabulut04}, Empty-Maximal-Space strategy~\cite{ramos16}, and Heightmap Minimization strategy~\cite{hu17, wang19_1}.

The first work to apply RL to solve the packing problem is Hu et al.~\shortcite{hu17}, which uses the framework proposed in~\cite{bello17} to find the packing order and further allows each object to have six different orientations when searching for the optimal packing. The more recent work along this direction is Duan et al.~\shortcite{duan19}, which proposes a new multi-task selection learning approach to learn a heuristic-like policy which generates the sequence and orientations of items to be packed simultaneously. 
%

However, none of the methods in the previous works can be directly applied to solve the TAP problem since the initial packing state provides strong constraints on the packing order and orientations of the objects. Also importantly, all the previous methods assumed that the input objects were provided without any extra spatial or dependency constraints, such as the ones represented using the precedence graphs in our work.


\paragraph{Neural Combinatorial Optimization.}
One of the best known neural architectures for tackling combinatorial problems is Pointer Network (Ptr-Net)~\cite{vinyals15}. It is a supervised 
sequence-to-sequence (Seq2Seq)~\cite{sutskever2014sequence} model trained to produce approximate solutions to such combinatorial problems as planar 
convex hulls, Delaunay triangulations, and the planar Travelling Salesman Problem (TSP). 
Bello et al. \shortcite{bello17} incorporate RL into Ptr-Net, obtaining an unsupervised learning paradigm. They empirically demonstrate that for TSP,
the generalization capability of the supervised Ptr-Net falls behind that of an RL agent who explores different tours and observes their corresponding rewards.

Ptr-Net is not an ideal fit to the TAP problem since our input is a {\em set\/} rather than a sequence. Hence, instead of Seq2Seq, we seek a Set2Seq learning framework. The recent RL-based network by Nazari et al.~\shortcite{nazari18} for Vehical Routing Problem (VRP)~\cite{golden2008vehicle} falls under Set2Seq; it forgoes the RNN encoder used in Ptr-Nets, which may add extra and unnecessary complications to the encoder when there is no meaningful order in the input set. \rh{One commonality between their network and \tapnet{} is that both allow some elements of each input to change between the decoding steps. However, the dynamic information used in ~\cite{nazari18} is computed for each node individually while our dynamic precedence information is extracted for the whole set. Moreover, to make sure that a box can be selected with different orientations, different box stats must be encoded separately. As a result, how the input should be encoded and output should be rewarded are quite differently
between these two works. Finally, differently from~\cite{nazari18}, since object selection and packing highly depend on the current packing state in the target container, instead of only feeding the network with the selected object, we also pass the intermediate packing state after that object is packed as input to \tapnet{}.}

\section{Overview}
\label{sec:overview}

Figure~\ref{fig:overview} shows an overview of our learning-based transport-and-pack framework, illustrated using a 2D example.
Given an initial spatial configuration of boxes, our goal is to transport and pack the boxes compactly into a target container.
Our method starts by analyzing the packing precedence among the boxes. 
The analysis result is expressed by a {\em precedence graph\/}, where each box corresponds to a graph node and the different 
precedence relations between the boxes are indicated by directed edges; see Figure~\ref{fig:overview}(b). 

\begin{figure}[!t]
    \centering
    \includegraphics[width=\linewidth]{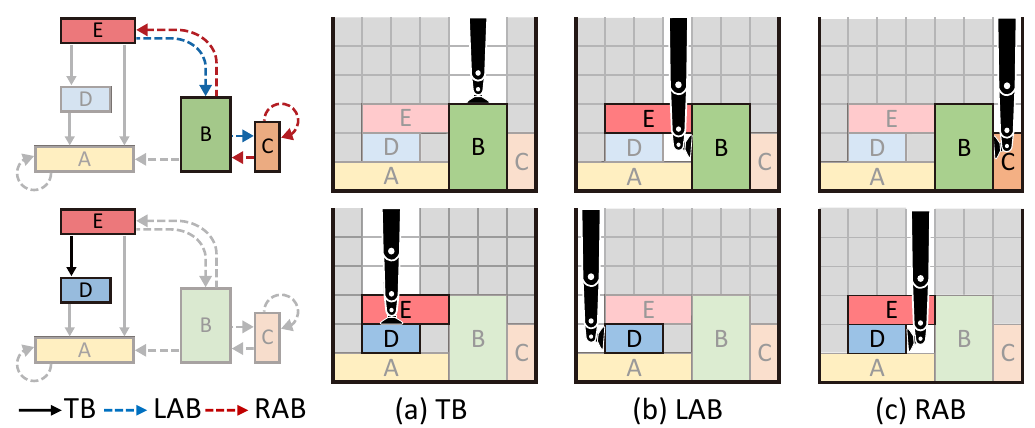}
\caption{
Top and side blocking illustrated in 2D. (a) TB: top blocking of box D by E, shown as black solid directed edge in precedence graph (left). (b) LAB: left access block of box B by E. (c) RAB: right access block of box B by C.}
\label{fig:access}
\end{figure}

There are three types of edges characterizing different precedence types; see Figure~\ref{fig:access}. A Top Block (TB) edge from $R$ to $O$ indicates that $R$ is on top of $O$ so that $O$ cannot be transported until $R$ has been packed. A Left Access 
Block (LAB) or Right Access Block (RAB) edge from $R$ to $O$ implies that $R$ blocks the access to the left or, respectively, the right side of $O$, which limits the rotation of $O$.
A special case of LAB and RAB
is when the initial set of boxes were found in a container and the container boundaries would block the access. In such a case, we use a self-loop LAB/RAB edge in the graph. More discussions on the precedence graph are given in Section~\ref{sec:precedence}.

Once the precedence graph is constructed, it is fed into our network \tapnet, together with the current box packing status of the target container, represented as a {\em height map\/}. The network is trained to output an optimal {\em packing order\/} and the associated box 
orientations. Note that the same box under different rotations are considered as different states and the number of states depends on the dimensionality of the boxes: six states for 3D boxes and two for 2D boxes. As a result, for a total of $N$ boxes to pack, \tapnet{} needs to choose among $2N$ or 
$6N$ states at each step. 
Once a box and its orientation are chosen, how it is packed is determined using a packing strategy, which 
we discuss in Section~\ref{sec:packing}.

The \tapnet{} architecture consists of an asymetric encoder-decoder pair, where the 
embedding or encoding layer encodes the input into a high-dimensional vector space. The decoder outputs the packing order
box orientations through an RNN with an attention mechanism. Specifically, at each step 
of the decoder, the embedded input is combined with an attention-weighted version as input to the RNN module to 
predict the next action, where the attention weights are computed from the embedded input and the RNN's hidden
state. The network weights used for embedding, decoding, and the attention layer are trained using policy gradients
with reward defined by packing efficiency and stability in the target container.

Information about the selected box is also used to update both the precendence graph and the height map characterizing the current packing state
of the container. Subsequently, the updated graph and height map are employed for our trained network TAP-Net to select the next box to pack and 
its orientation. This process repeats untill all the input boxes are packed into the target container.

\section{Method}
\label{sec:method}

Here we start with introducing how to extract precedence constraints based on the initial packing state in Subsection~\ref{sec:precedence}. The network architecture and how to train the \tapnet{} for optimizing the order and orientations of box packing are then discussed in Subsection~\ref{sec:architecture} and~\ref{sec:train}. Once a box is chosen, different packing placement methods can be used for placing it in the target container, which are explained in Subsection~\ref{sec:packing}. For simplicity, the above methods are discussed using 2D boxes. We further assume that all boxes are packed to the same target container and the total number of boxes is smaller than or equal to the capacity of the trained network. How to use the trained network to handle larger instance set, how to pack objects into multiple containers, and how to extend the algorithm to 3D cases are explain in Subsections~\ref{sec:extend_larger}, \ref{sec:extend_multi}, and \ref{sec:extend_3D}, respectively.

\subsection{Precedence extraction}
\label{sec:precedence}

\begin{figure}[!t]
    \centering
    \includegraphics[width=0.48\textwidth]{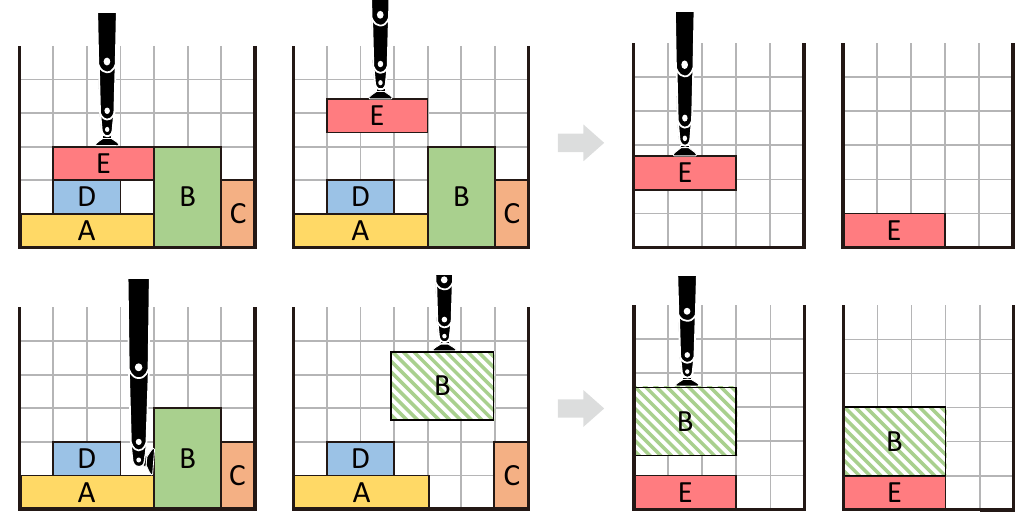}
\caption{
Illustration on how to transport-and-pack objects without (top) and with (bottom) rotation. Left side shows the original object configuration, whereas right side shows the target container.}
\label{fig:tap_example}
\end{figure}

As discussed above, our approach aims to address a variant of physical packing problem in which not all objects are ready to be packed into the target container at the same time. For example, when the objects are initially packed inside another storage, we need to select and pack objects on top before handling those underneath.  Hence, the accessibility of a given object depends on all objects on top of it.  In addition, if we want to rotate an object, we assume that it can only be done through attaching the robotic arm to the side of the object first, lifting the object to an open space for rotation, and finally packing the object into the target container from the top; see Figure~\ref{fig:tap_example}. Under this assumption, an object can only be rotated when its side is accessible to the robotic arm.

Specifically, to find the precedence set of each box $O$, we need to locate the set of boxes that have to be removed before $O$ can be packed with or without rotation. For a 2D box, we need to check its top, left, and right sides. For example, box $E$ in Figure~\ref{fig:access} needs to be moved first to allow the top side of box $D$ to be accessible.  Hence, a TB edge connects from node $E$ to node $D$. There isn't any TB edge pointing to node $B$ since there is no object on top of it. The accessibility of the left side of box $B$ is blocked by box $E$ and hence a LAB edge is added from $E$ to $B$. Similarly, an RAB edge is added from $C$ to $B$ as box $C$ blocks the access to right side of $B$.
It is worth noting that, when the left (or right) side of a box is attaching to the container, the box is considered being blocked from the left (or right) side. This is represented using a LAB (or RAB) edge linking to itself; see the boxes $A$ and $C$ as examples.

To pack any box $O$ in its original orientation, the only condition is that there is no TB edge pointing to it. However, to pack $O$ under a rotated state, we need to further require its left or right side to be accessible.  Please note that here we assume all objects can be rotated both clockwise and counterclockwise. Hence the rotation is enabled when either side is accessible.  To minimize the number of edges in the graph, when the left (or right) side of an object is already accessible, we will not add any RAB (or LAB) edge to it; see node $D$ in Figure~\ref{fig:access} for example.

\subsection{\tapnet{} architecture}
\label{sec:architecture}

Figure~\ref{fig:network} shows the network architecture of our \tapnet{} on 2D cases. 
Similar to previous sequence generation models, \tapnet{} consists of an encoder and a decoder with an attention mechanism. The two modules collectively determine a sequence of optimal states $\{y^t\}_{t=1,\dots,n}$, which is used to pack boxes into the target container. Note that under 2D, each box has two different states: original and rotated orientations. Hence, $n$ boxes result in $2n$ states.  The network only outputs $n$ states though, as only one of the two states can be selected for each box.

\begin{figure}[!t]
    \centering
    \includegraphics[width=0.48\textwidth]{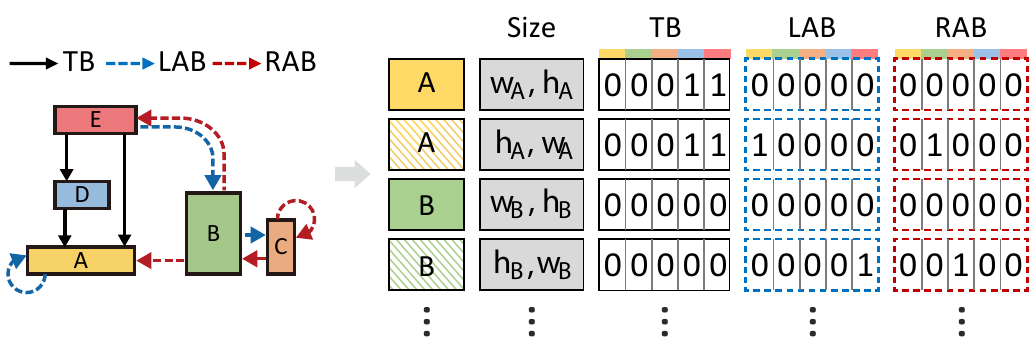}
\caption{
The encoding of box states. Each box is encoded under two states (rows) for original (solid) and rotated (stripe pattern) orientations. Besides static width and height information, each state also records the dynamic dependency information using binary codes. Each bit in the binary code represents one of the objects. For example, Object A is blocked by objects $D$ and $E$ from the top and hence the values at the corresponding bits in TB are set to `1'.} 
\label{fig:encoder}
\end{figure}

\begin{figure}[!t]
    \centering
    \includegraphics[width=0.48\textwidth]{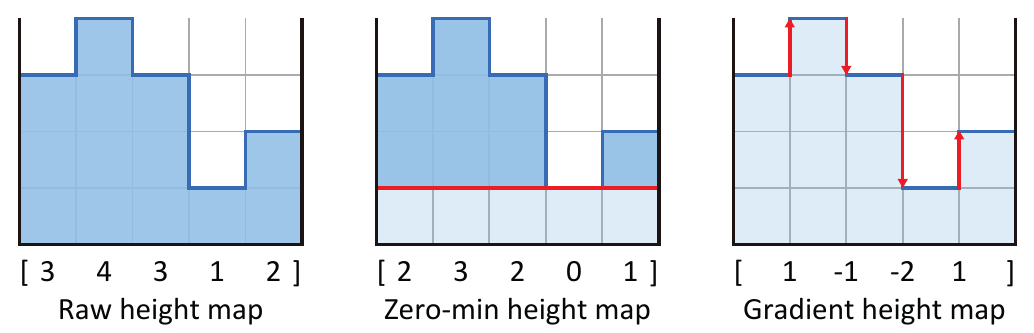}
\caption{
\rh{Three different height map representations under the same target container configuration.}
}
\label{fig:heightmap}
\end{figure}

The input for the encoder module is denoted as $\{s_i, d_i^{0}\}$, where $s_i = \{w_i, h_i\}_{i=1,\dots,2n}$ represents the static width and height information and $d_i^t$ represents the dynamic dependency set of the $i^{th}$ box state. Figure~\ref{fig:encoder} shows how the information is encoded for each object orientation in 2D. The decoder then takes the previously selected state $y^{t-1}$ as input. In addition, to make the network aware of the current status inside the target container and hence be able to adaptively select the next state, we also feed the dynamic height map of the container $h^{t-1}$ into the decoder.
\rh{Three different height map representations are considered. The raw map directly uses the height values inside the container. Considering that we are more interested in height variation than absolute heights, we also tested zero-min and gradient height maps; as shown in Figure~\ref{fig:heightmap}. The impact of height map representations on algorithm performance is discussed in Sec.~\ref{sec:data}.}

For each step $t$,  attention mechanism is used to accumulate information till step $t-1$ and output the probability map over all valid states. The one with the maximum probability is selected to be the output $y^t$ for each step $t$; see Figure~\ref{fig:network}. 
In this example, the first state $s_1$ is selected as output $y^t$. Once a state $y^t$ is picked, the following operations are performed: i) the box $O$ and its orientation represented by $y^t$ is packed into the target container using selected packing strategy; ii) the precedence graph is updated by removing the corresponding node and all connected edges; iii) the container height map $h$ is updated to reflect changes in the target container; and iv) since $O$ has already been packed, the two states associated to $O$ are no longer valid choices.
To label them, a masking scheme is used, which sets the log-probabilities of invalid states to $-\infty$.

Once the above four operations are performed, the updated precedence information and container height map are fed into the network for outputting the optimal state $y^{t+1}$ for the next step $t+1$. The process iterates until a sequence of $n$ states is obtained.



\begin{figure}[!t]
    \centering
    \includegraphics[width=0.48\textwidth]{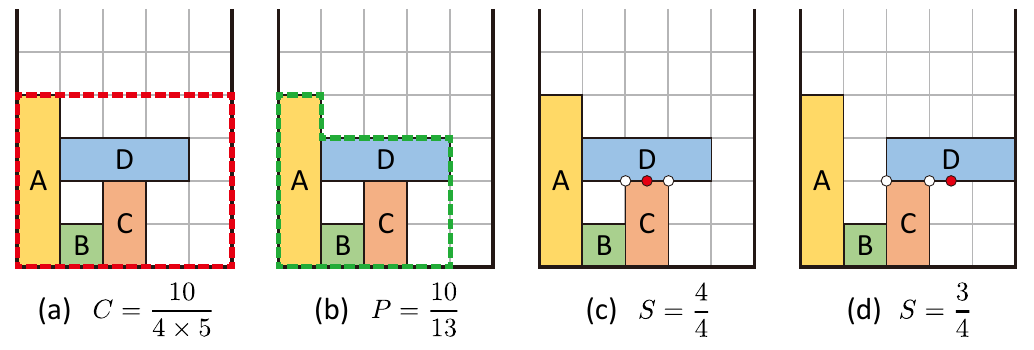}
\caption{
Packing reward calculation. (a) Compactness $C$ is computed using the ratio between the total area of all shapes $A_{\text{shape}}$ and the rectangular area defined by the maximum height (dashed red line). (b) Pyramidality $P$ is defined as the ratio between $A_{\text{shape}}$ and the area of the region obtained by projecting all shapes toward the bottom of the container (dashed green line).  (c-d) Stability $S$ is measured as the percentage of stable shapes among all shapes packed so far. To determine whether shape $D$ is stable, we first locate all its supporting points (white dots). If $D$'s center falls inside the region spanned by the supporting points, it is considered as stable (c); otherwise it is not (d).}
\label{fig:reward}
\end{figure}

\subsection{Network training}
\label{sec:train}
To train the network, we use the well-known policy gradient approaches~\cite{mnih2016asynchronous, bello17}. 
The training method is quite standard for RL, so we leave the details in the supplementary materials and only discuss the packing reward here. 

To measure the packing quality, we define the reward function as:
\begin{equation}
R = (C+P+S) / 3,
\label{eq:reward}
\end{equation}	
where $C$ is the Compactness, $P$ is the Pyramidality, and $S$ is the Stability of the packing. 

More specifically, the compactness $C$ is defined as the ratio between the total area of packed boxes $A_{\text{packed}}$ and the minimum container size needed.  The latter is computed as the rectangular area $A_{\text{rect}}$ specified by the maximum packing height and container width $W$ as shown in Figure~\ref{fig:reward}(a).  Intuitively, the compactness measure favors tightly packed boxes and $C=1$ when all boxes fully fill the container up to a given height.

The pyramidality $P$ is defined as the ratio between $A_{\text{packed}}$ and the area of the region obtained by projecting all boxes toward the bottom of the container $A_{\text{proj}}$; see Figure~\ref{fig:reward}(b). Since the area outside of projected region can be filled by future objects, the pyramidality measure outputs high values when there is good potential to tightly pack all objects.

Finally, the stability $S$ is defined as the percentage of stable boxes $N_{\text{stable}}$ among all boxes packed so far $N_{\text{packed}}$. To determine whether a given box $D$ is stable, we adopted the method used in \cite{ramos16}, which first locates all $D$'s supporting points. If and only if $D$'s center falls inside the region spanned by these supporting points, it is considered as stable; see Figure~\ref{fig:reward}(c) and (d).
Please note that we do not strictly enforce that an object needs to stable when packing, as we assume filling materials can be used for additional support. Nevertheless, stable packing is preferred and hence we favor solutions with high $S$ values.

\subsection{Packing placement methods}
\label{sec:packing}

How well the objects can be packed in the target container depends on both the order used for packing these objects and where each object is placed in the container.  Since optimizing both \change{the packing order and placement} simultaneously could be an intractable task \change{due to the huge and mixed discrete-continuous searching space}, heuristically designed strategies are used for object placement in existing approachs~\cite{hu17, duan19}.  These include Deepest-Bottom-Left with Fill (DBLF) strategy~\cite{karabulut04} and those based on the Empty-Maximal-Space (EMS) \cite{ramos16}. Two EMS-based packing placement strategies are integrated and tested in our approach.  Both use the EMS found in the remaining space of the target container. Their key difference is how to select the candidate position inside each EMS. The first strategy, denoted as LB, only tries the bottom-left corner of each EMS and selects the one with the highest packing reward. The second strategy, denoted as MUL \change{(Multiple EMS)}, tests all four bottom corners of each EMS and uses the same packing reward to select the optimal candidate. 

Regardless how a given number of boxes are packed, the amount of empty space inside the target container is fixed.  However, due to accessibility, not all empty spaces can be utilized in future packing. In addition, to accommodate large objects, a single connected empty space is preferred over multiple separated spaces. 
Motivated by this finding, we also designed a new heuristic strategy that optimizes the empty space available for packing future, potentially large objects, which is referred to as Maximize-Accessible-Convex-Space (MACS) strategy.
Here we measure the amount of \emph{accessible convex space}, as we consider only convex space is usable for placing a single large object. 
When placing a new box $O$, our strategy is to maximize the accessible convex space after $O$ is positioned.  Specifically, we first compute all the candidate positions for placing $O$ through locating all the corner locations at different layers.  Box $O$ is then placed at each of these candidate positions to evaluate the amount of usable convex space afterward. The candidate location that yields the maximum remaining usable space is chosen. 

\begin{figure}[!t]
    \centering
    \includegraphics[width=0.48\textwidth]{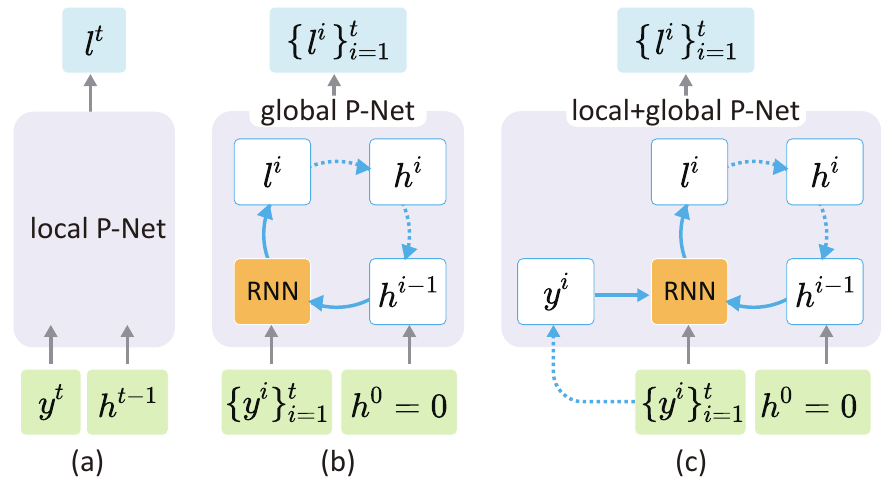}
\caption{
\rh{Network structures for three packing placement networks. When a new object $y^t$ is introduced, local P-Net (a) only outputs its placement location $l^t$. Global P-Net (b) and local+global P-Net (c) recompute the placement locations for all objects in the container and hence output $\{l^i\}^t_{i=1}$. } 
}
\label{fig:p-net}
\end{figure}

\begin{figure}[!t]
    \centering
    \includegraphics[width=0.49\textwidth]{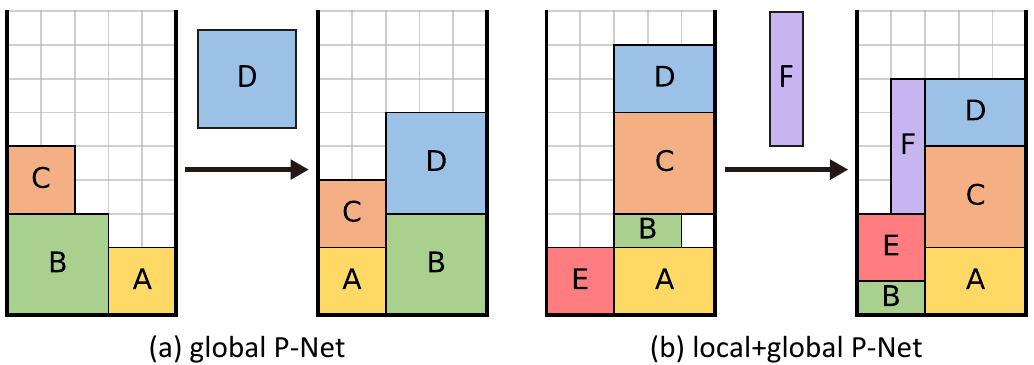}
\caption{
\rh{Example results of global P-Net (a) and  local+global P-Net (b), where the packing locations of previous objects are updated when a new object given to pack.}
}

\label{fig:lg-pnet}
\end{figure}

\begin{figure*}[!t]
    \centering
    \includegraphics[width=\textwidth]{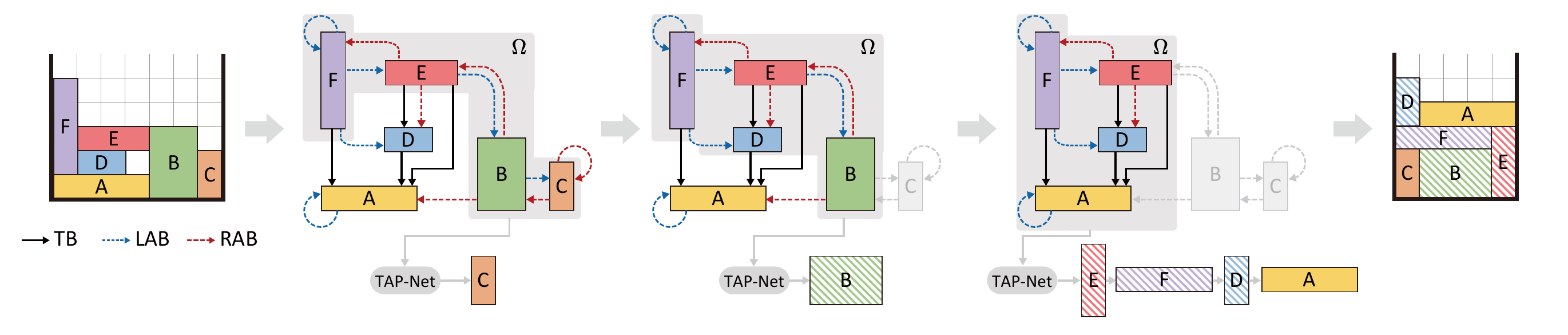}
\caption{
Rolling-based packing for handling large instance set. When TAP-Net is trained for 4 objects and there are 6 objects to be packed, we first select 4 objects into the initial set $\Omega$, based on which TAP-Net chose the first object to be packed ($C$ in this case).  After $C$ is packed and removed, another object ($F$ in this case) is added to $\Omega$.  The process repeats until all objects are packed.}
\label{fig:rolling}
\end{figure*}

\rh{
Besides the above heuristic packing placement strategies, three learning-based approaches are also designed. All three networks are trained to determine where to place an input object of size $(w_i, h_i)$ by outputting a horizontal position for the object's left edge $l^t$. If $l^t+w_i > W$, we set $l^t = W-w_i$ so that the object can fit inside the width $W$ container. Once $l^t$ is determined, we let the object drop from the top to determine its vertical location. The height map of the container is then updated accordingly.

A key difference among the three networks lie on their inputs; see Figure~\ref{fig:p-net}. The first network takes the current container height map and packing object $y^t$ as input and output $l^t$ for object $t$ only. Since the decision is made for one object at a time, it is referred to as local P-Net. The second network takes the whole sequence of objects that have been packed $\{y^i\}^t_{i=1}$ as input and uses RNN to recompute placement locations for each object sequentially, i.e., $\{l^i\}_{i=1}^t$. We here call it global P-Net.  Similar to global P-Net, the third network takes $\{y^i\}^t_{i=1}$ as input and hence uses global information.  It also takes the new object $y^i$
as additional input, which draws attention to the current state.  Hence, it is termed local+global P-Net.

It is worth noting that every time a new object is added, global P-Net and local+global P-Net regenerate the placing locations for all objects in the container. This helps to optimize packing results (see Figure~\ref{fig:lg-pnet}) and is a feature not available in local P-Net, LB, MUL, and MACS methods.
All three networks are trained using reinforcement learning, with the aforementioned packing reward $R$ serving as the objective function.
When being used with the proposed \tapnet{}, we tried both end-to-end training from scratch and pre-train ordering/placing networks separately followed by fine-tuning, and more details can be found in the supplementary materials.

\change{We present the comparisons of different packing placement methods in Section~\ref{sec:baseline}, and it shows that heuristic packing strategies yield better results and we adopt LB strategy in our method due to its simplicity.}
}

\subsection{Extension to larger instance set}
\label{sec:extend_larger}

\change{As other RL models}, when training \tapnet{}, we need to predetermine the capacity of the network in terms of the number of objects $n$. This is because the dimension of the input vector $d_i$, as well as the dimension of output feature vectors of both encoder and decoder, depend on value $n$. Once the network is trained, it can deal with input with smaller instance set through filling the remaining entries with dummy data. However, handling larger set of input objects would normally require retraining the network.

On the other hand, due to obstruction and
accessibility constraints, when there are $m$ $(m>n)$ objects in the initial configuration, not all of them are ready to be packed.  Hence, we can choose a set ($\Omega$) of $n$ objects for \tapnet{} to transport and pack. Once an object has been packed, it will be removed and another object will be added to $\Omega$. Figure~\ref{fig:rolling} shows an example with \tapnet{} trainined on 4 objects and applied on 6 objects, which shows how \tapnet{} operates on 4 objects at a time.  We refer this strategy as rolling-based packing.

When choosing $n$ objects for the initial set $\Omega$, we want to pick the ones that are ready to be packed or will soon be ready. Given the initial box configuration, there is always at least one object that is accessible from top and can be packed right away.  We give these objects the highest priority for placing into $\Omega$. Any other object $O$ will have its priority determined by the minimum number of objects that have to be removed before $O$ can be accessed from the top. In addition, between two objects that have the same priority value, the one that can be accessed with rotation is put into set $\Omega$ first.


In the example shown in Figure~\ref{fig:rolling}, the first four boxes selected into set $\Omega$ based on the rules above are $F$, $E$, $B$, and $C$. These four boxes are fed into \tapnet{}, which picks box $C$ for packing first. Box $D$ is then selected from the remaining precedence graph, which replaces $C$ as part of the input to network. Note that once $C$ is packed, all the precedence edges starting from $C$ will be deleted and the corresponding constraint on the rotations may be removed as well. For example, packing $C$ allows object $B$ to be packed with rotation. The process iterates until all the objects are packed, and the final packing result is shown on the right.

\subsection{Extension to multi-containers}
\label{sec:extend_multi}

\begin{figure}[!t]
    \centering
    \includegraphics[width=0.48\textwidth]{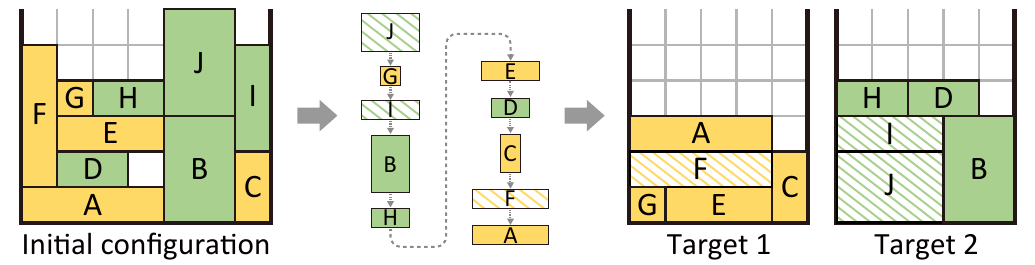}
\caption{
Extension for packing into multiple containers. Objects in the initial container need to be packed into two different target containers, yellow ones to Target 1 and green ones to Target 2. Our approach is able to generate an optimal packing sequence in this case as well.}
\label{fig:multi}
\end{figure}

The \tapnet{} discussed so far assumes that all objects are to be packed into the same target container. In some real-world applications, however, objects may need to be packed into different target containers. For example, in delivery industry, packages from the same container may need to be shipped to different customers or distribution centers. In these cases, which of the target container that a given object should be packed into is known.

To extend \tapnet{} for handling cases of packing boxes into $k$ target container, two major changes are needed. First, we need to add the target container index $idx \in [1, k]$ to the static input of each object when encoding.  Secondly, the height maps of all target containers $\{h_j\}_{j=1,2, \dots, k}$ need to be fed into the decoder, which helps the network to select which object and the corresponding target container should be packed next.

Figure~\ref{fig:multi} shows an example of transporting object from one container to two target containers.  The target container for each object, which is assumed to be given as part of the input, is indicated by different colors.


\subsection{Extension to 3D}
\label{sec:extend_3D}
As mentioned above, we chose to explain the algorithm in 2D first for simplicity. Now we discuss all the changes needed for handling 3D cases.

\begin{figure}[!t]
    \centering
    \includegraphics[width=0.44\textwidth]{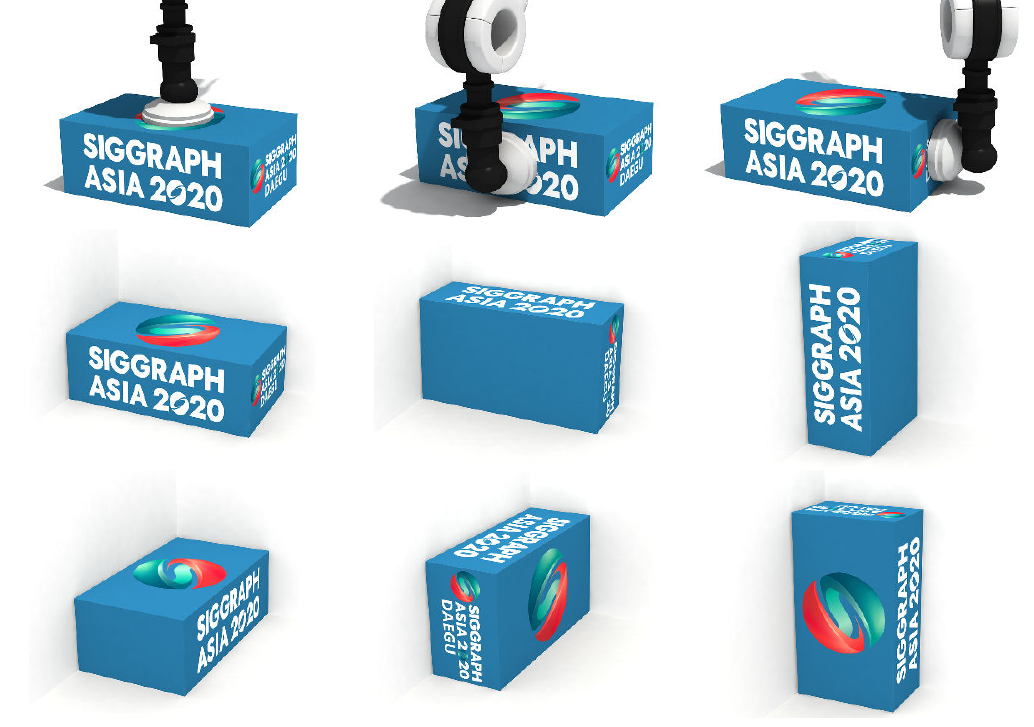}
\caption{Six different object rotation states in 3D (bottom 2 rows) and the corresponding robot arm access approach (top row). The first two states (left column) only require the top of the box being accessible, whereas the remaining four also require one of the sides being accessible.}
\label{fig:rotation_3D}
\end{figure}

First of all, with rotation enabled, a 2D shape has only two different states. A 3D object, however, has six different states resulting from rotating around different axes; see Figure~\ref{fig:rotation_3D}. Hence, the number of input states for $n$ objects becomes $6n$, instead of $2n$. The height map $h$ of the target container also becomes a 2D image rather a 1D vector.

\begin{figure}[!t]
    \centering
    \includegraphics[width=0.44\textwidth]{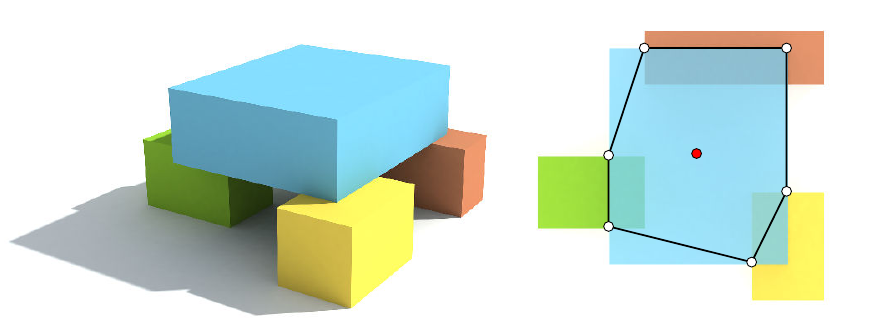}
\caption{
3D stability evaluation. An object is considered as stable if the projection of its center is inside the region formed by supporting points.}
\label{fig:stable}
\end{figure}

Moreover, all packing reward calculations need to be adapted to 3D. The extension of compactness and pyramidality to 3D is relatively straightforward.  Determining whether an 3D object is stable is more difficult than 2D.  As shown in Figure~\ref{fig:stable}, we first need to intersect the 2D profile of the 3D object with those of its supporting objects. This gives us a set of supporting points. The object is considered as stable if its center falls inside the convex hull of these supporting points.

\section{Results and evaluation}
\label{sec:results}

To train the network, we only need to prepare a large number of initial box configurations, and rely on the reward functions and policy gradients under the classical RL framework to learn optimal TAP policies. There is no need for data labeling or feeding the optimal packing solutions for input data into the network. Without loss of generality, we here train TAP-Net to handle 10 objects and the target container has width $W=5$ and unlimited height.

\subsection{Data preparation and representation}
\label{sec:data}

Two sets of data are prepared in our approach. The first set of initial box configurations were generated randomly, which we call the \emph{RAND} set. Specifically, the \emph{RAND} set contains 120K training samples and 10K testing samples. To generate each data sample, we first randomly generate 10 boxes, where the width and height of each box is sampled from a discrete Gaussian distribution between 1 and 5. The initial configuration of these 10 boxes are obtained through packing them into a container ($W=7$) using a random order and each box under a random orientation. It is worth noting that the container width setting here affects how likely different boxes block each other. That is, when the width of the container is smaller, boxes are more likely to be on top of each other, and the precedence graph extracted from the initial configuration will likely have more edges.

Since each initial box configuration in the \emph{RAND} set is randomly generated, how compact the boxes can be packed into the target container under the optimal solution is unknown, making it difficult to evaluate how close a given solution is to the optimal solution.  To address this problem, the second data set, termed \emph{Perfect Packing Solution Guaranteed} (\emph{PPSG}) set, is also generated. Here each initial box configuration is derived from a {\em randomly generated but perfectly packed\/} target container, which is referred to as a Candidate Perfect Packing Solution (CPPS). We take the boxes out of the target container and pile them up  \change{in sequence} to form an initial configuration, which guarantees to be reversible for packing back into the CPPS.

Specifically, we first choose a $5\times H$ block as the result of perfectly packing 10 boxes, where $H$ is computed based on the distribution of total box sizes in the \emph{RAND} set.  We then split this $5\times H$ rectangle block into 10 boxes. This is achieved through iteratively picking and splitting one of the existing block (starting with only one) into two smaller ones, until we have all 10. The probability of a given block being picked is proportional to its area, the probability of cut direction is proportional to the dimension of the block perpendicular to that direction, and the  probability of the cutting location is proportional to its distance to the center of the block. 
	
\begin{figure}[!t]
    \centering
    \includegraphics[width=0.49\textwidth]{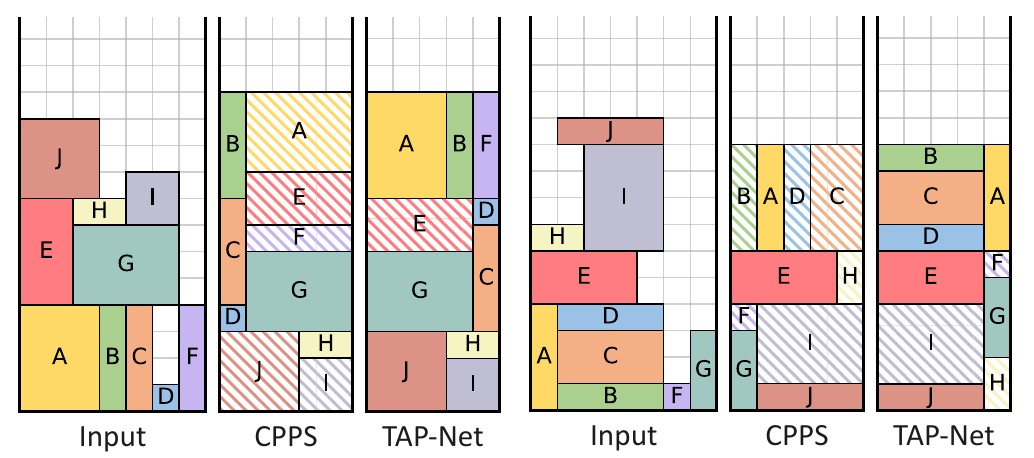}
\caption{
Each initial box configuration (two are shown here) in the PPSG set is generated from a Candidate Perfect Packing Solution (CPPS). CPPS is not used for training and \tapnet{} often finds optimal transport-and-pack solutions that are different from CPPS.}
\label{fig:comp_gt}
\end{figure}

Once the $5\times H$ rectangle block is split into 10 boxes, we have a CPPS for these boxes.  Next, we transport and pack them into a container with $W=7$ to form an initial configuration; see Figure~\ref{fig:comp_gt}. Same as in \emph{RAND} set, here we also randomly rotate each box to add variation to the data. However, for each rotated box, we only place it to a position where either its left or its right side is accessible.  This is to make sure that the box can be accessed under rotation state when packing it back to the target container in reverse order.

Since we guarantee each data sample in \emph{PPSG} can be tightly packed back into a rectangle area, the $C$, $P$, and $S$ measures of its optimal packing solution all equal to 1. It is worth noting that the CPPS used for generating the initial box configuration is not provided to \tapnet{} for training. In fact, \tapnet{} often outputs an optimal packing solution that is different from the initial CPPS; see Figure~\ref{fig:comp_gt}.

\rh{We also quantitatively evaluated the performance of \tapnet{} on both \emph{RAND} and \emph{PPSG} datasets and using each of the three height map representations discussed in Sec.~\ref{sec:architecture}; see Table~\ref{tab:comp_heightmap_type}. As expected, under the same settings, the $C$, $P$, and $S$ measures on \emph{PPSG} dataset are higher since data samples in this set can be densely packed.  
In addition, using \emph{Gradient} high map representation consistently yields best performance. Hence, \emph{Gradient} height map is used in all remaining experiments.}

\begin{table}[!t]%
	\caption{\rh{The performance of \tapnet{} under different height map represenations on both RAND and PPSG datasets.}}
	\label{tab:comp_heightmap_type}
	\begin{minipage}{\columnwidth}
		\begin{center}
		\begin{tabular}{l||l||l|l|l|l}
			
            \hline\hline
			\textbf{Data} & \textbf{Height map} &   \bm{$C$} & \bm{$P$}&  \bm{$S$} &  \bm{$R$}   \\ \hline \hline 
			\multirow{3}{*}{RAND} 
            & Raw & 0.910 & 0.984 & 0.990 & 0.961 \\ \cline{2-6} 
			& Zero-min & 0.915 & 0.986 & 0.995 & 0.965 \\ \cline{2-6} 
			& \textbf{Gradient} & \textbf{0.925} & \textbf{0.988} & \textbf{0.997} & \textbf{0.970} \\ \hline \hline 
			
            \multirow{3}{*}{PPSG} 
            & Raw & 0.970 & 0.988 & 0.996 & 0.985 \\ \cline{2-6} 
			& Zero-min & 0.959 & 0.986 & 0.996 & 0.980  \\ \cline{2-6} 
			& \textbf{Gradient} & \textbf{0.981} & \textbf{0.996} & \textbf{0.999} & \textbf{0.992} \\ \hline \hline

		\end{tabular}
		\end{center}
	\end{minipage}
\end{table}%

\rh{
\subsection{Baseline algorithms \rh{and comparisons}}
\label{sec:baseline}

Given that transport-and-pack is a brand new problem and there is no existing approach to compare with, we implemented a couple of baseline algorithms. To focus the evaluation on the packing order generated by \tapnet{}, all baseline algorithms use the same precedence graph generation procedure and the same set of packing placement methods as \tapnet{}.

The first baseline algorithm, referred to as \emph{Random}, selects randomly among nodes in the precedence graph that have zero in-degree at each step.  Once the corresponding box under the orientation associated with the node is placed in the target container based on the packing strategy, the precedence graph is updated by removing related nodes and edges.


The second baseline approach tries to locally optimize node selection at each step, rather than randomly picking among them. That is, all nodes with zero in-degree \change{in terms of all kinds of edges} are enumerated. A packing reward is calculated after placing each of the corresponding box in the target container. The node that leads to the highest packing reward is selected. As expected, this method, referred \emph{Greedy}, yields better performance than \emph{Random}. However, it doesn't scale well and is time-consuming when the number of objects increases.

\paragraph{Comparison with baseline algorithms under different packing placement methods}
As discussed above, there are two key components for solving the TAP problem: choosing the order (including orientations) for packing different objects and deciding where to place each object.  \tapnet{} only optimizes packing order and orientations used for packing. Where to place each object in the target container is determined by packing strategies. Six different packing strategies have been discussed in Sec.~\ref{sec:packing}. Among them, two existing (LB and MUL) and one developed (MACS) strategies are rule-based, whereas the other three are learning-based. To thoroughly compare the performances of \tapnet{} with the two baseline approaches, we quantitatively evaluated all $3\times 6$ combinations on the RAND dataset; see Table~\ref{tab:comp_packing}.


\begin{table}[!t]%
	\caption{\rh{Performances of different packing ordering and placement methods on RAND dataset. While the Random and Greedy ordering approaches perform the best under MACS placement method, the proposed \tapnet{} algorithm works equally well under all three heuristically designed packing placement methods.}}
	\label{tab:comp_packing}
	\begin{minipage}{\columnwidth}
		\begin{center}
			\begin{tabular}{l||l||l|l|l|l}
				\hline\hline
				\textbf{Order} & \textbf{Packing} &   \bm{$C$} & \bm{$P$}&  \bm{$S$} &  \bm{$R$}   \\ \hline \hline 
				
				
				\multirow{6}{*}{Random} 
				& LB  & 0.736 & 0.898 & 0.946 & 0.860  \\  \cline{2-6} 
				& MUL & 0.735 & 0.892 & 0.943 & 0.857  \\  \cline{2-6} 
				& \textbf{MACS} & 0.747 & \textbf{0.901} & \textbf{0.946} & \textbf{0.865} \\\cline{2-6} 
				& L P-Net  & 0.523 & 0.772 & 0.900 & 0.731 \\ \cline{2-6} 
				& G P-Net & 0.669 & 0.833 & 0.890 & 0.797 \\ \cline{2-6} 
				& LG P-Net & \textbf{0.763} & 0.892 & 0.911 & 0.856 \\  \hline
				\hline
				\multirow{6}{*}{Greedy}
				& LB  & 0.816 & 0.972 & 0.977 & 0.922  \\  \cline{2-6} 
				& MUL & 0.815 & 0.972 & 0.978 & 0.922  \\  \cline{2-6} 
				& \textbf{MACS} & 0.816 & \textbf{0.972} & 0.978 & \textbf{0.922}  \\  \cline{2-6}
				& L P-Net  & 0.546 & 0.903 & \textbf{0.980} & 0.810 \\ \cline{2-6} 
				& G P-Net  & 0.698 & 0.904 & 0.945 & 0.849 \\ \cline{2-6} 
				& LG P-Net & \textbf{0.825} & 0.957 & 0.957 & 0.913 \\  \hline \hline
				
				\multirow{6}{*}{\tapnet} 
				& \textbf{LB}  & \textbf{0.925} & \textbf{0.988} & 0.997 & \textbf{0.970}  \\  \cline{2-6} 
				& MUL & 0.925 & 0.987 & 0.997 & 0.970  \\  \cline{2-6} 
				& MACS & 0.922 & 0.988 &  \textbf{0.998} & 0.969  \\  \cline{2-6} 
				& L P-Net  & 0.593 & 0.852 & 0.946 & 0.797 \\\cline{2-6} 
				& G P-Net & 0.819 & 0.937 & 0.973 & 0.910 \\ \cline{2-6} 
				& LG P-Net & 0.898 & 0.983 & 0.989 & 0.957 \\ \hline \hline
				
				
			\end{tabular}
		\end{center}
	\end{minipage}
\end{table}%

As shown in Table~\ref{tab:comp_packing}, under the Random packing orders, the presented MACS approach achieves notable improvement over existing packing strategies (LB and MUL) and also achieved highest overall packing reward $R$. When the Greedy packing orders is used, the advantage of MACS over LB and MUL becomes unremarkable. This suggests that packing objects in proper orders can reduce the impacts of different placement methods.
Among the three learning-based packing placement methods, the local+global P-Net (LG P-Net) achieves the best performance under both Random and Greedy packing orders. It also outperforms MACS in terms of Compactness $C$ in both cases.

Interesting results are observed when \tapnet{} is used to generate the packing order. All three heuristic packing strategies yield similar results and outperform the three learning-based methods. Our hypothesis is that the heuristic packing placement strategies are more \emph{predictable}, and hence it is easier for \tapnet{} to anticipate where the next box will be placed and output the optimal box and its orientation accordingly. 
While \tapnet{} is capable to adapt to three different heuristic placement strategies and provide nearly identical performance, it cannot work as well with learning-based placement methods since their outputs are harder to predict.

Despite the relative poor performance of \tapnet{} under learning-based placement methods, it actually outperforms both Random and Greedy ordering approaches by a large margin in all settings.  Since LB strategy is most efficient and works equally well as MUL and MACS, it is used in all the remaining experiments.

\paragraph{Performance of different approaches on the PPSG dataset.}
Table~\ref{tab:comp_baseline} further compares the performances of \tapnet{} and the two baseline approaches on the PPSG sets. For both baseline ordering approaches, MACS placement method is used since it yields best performance.  
The performances of \tapnet{} are very close to `1', whereas those of the two baseline approaches are much lower.
Compared to the performance of \tapnet{}+LB shown in Table~\ref{tab:comp_packing}, $C$ measure achieved on the PPSG dataset is noticeably higher than the one on RAND dataset (0.981 vs 0.925).

Comparison between the two baseline approaches show that Greedy ordering performs notably better than Random.  However, enumerating all candidate states introduces high computational cost, with the processing time increasing by more than four folds.  
The \tapnet{}, on the other hand, runs even faster than Random ordering MACS placement. This is mainly because \tapnet{} uses the faster LB placement method, whereas Random uses MACS. 

\begin{table}[!t]%
	\caption{\rh{Performance comparison between \tapnet{} and the two baseline ordering algorithms on PPSG dataset.}} 
	\label{tab:comp_baseline}
	\begin{minipage}{\columnwidth}
		\begin{center}
		\begin{tabular}{l||l|l|l|l|l}
			\hline\hline
			\textbf{Method} &   \bm{$C$} & \bm{$P$}&  \bm{$S$} &  \bm{$R$}  &  \bm{$t$}(ms) \\ \hline \hline 
			
			
			Random  & 0.771 & 0.849 & 0.911 & 0.843 &  8 \\  \hline 
			Greedy & 0.920 & 0.971 & 0.990 & 0.960 & 45 \\  \hline 
			\textbf{\tapnet{}} & \textbf{0.981} & \textbf{0.996} & \textbf{0.999} & \textbf{0.992} & \textbf{3} \\ \hline \hline 
		\end{tabular}
		\end{center}
	\end{minipage}
\end{table}%

\begin{figure}[!t]
    \centering
    \includegraphics[width=0.48\textwidth]{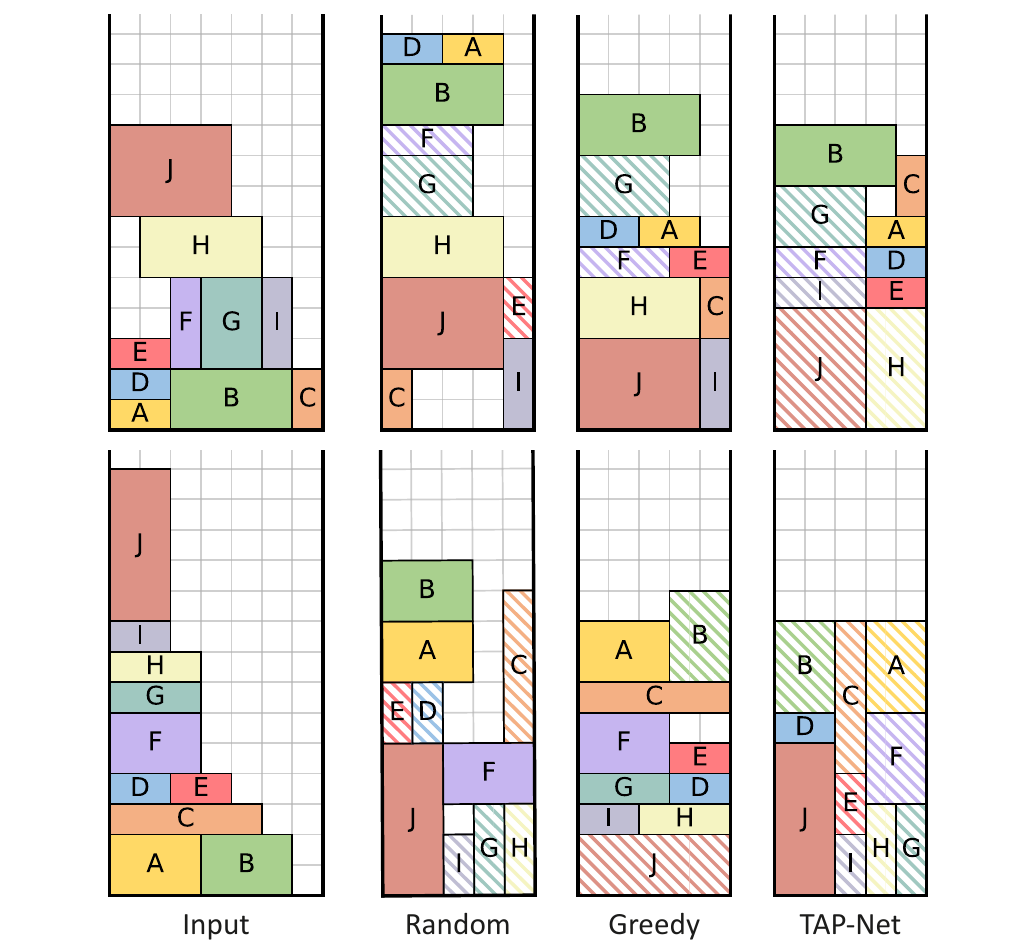}
\caption{ \rh{ Packing comparison on two data samples: one from the RAND set (top) and one from the PPSG set (bottom). \tapnet{} tightly packs boxes in both tests, whereas the Random and Greedy ordering failed to do so.}} 
\label{fig:comp_baseline}
\end{figure}

Figure~\ref{fig:comp_baseline} shows the comparison of some packing results. Compared to the Random ordering baseline, our method successfully learns a good transport-and-packing strategy to ensure an efficient and stable packing,  whereas Random ordering fails to form a compact packing, even under a more sophisticated packing placement method. For the baseline which searches the best node among all feasible ones, it does get better results than the baseline with randomness,  however, it cannot reach the similar performance obtained using \tapnet{}. Moreover, this kind of exhaustive search is time consuming comparing to other methods and cannot scale well when the number of objects increases.
}

\subsection{Ablation studies and parameter impacts}

\rh{
\paragraph{Ablation study on inputs to encoder}
Given the initial configuration of objects, \tapnet{} optimizes the packing order and object orientations using a Set2Seq network, which we designed for this task.  An alternative approach is to employ the representative Seq2Seq network, Ptr-Net~\cite{vinyals15}, which only accepts randomized static information from a sequence as input; see Table~\ref{tab:comp_ptr}. As an ablation study, we here compare the performances of our original \tapnet{}, \tapnet{} using only static information, and Ptr-Net using the same static information.  Two settings on available static information are tested, one with only sizes of different objects available (\emph{shape only}) and the other having additional information on initial precedence (\emph{shape + init P}), but without dynamically updating the precedence information.
	
The results in the table suggest that, under both settings on available static information, \tapnet{} performs better than Ptr-Net. This means that the Set2Seq network we use is more suitable to our problem than Seq2Seq networks. It is also worth noting that adding static initial precedence information does not improve either \tapnet{} nor Ptr-Net approaches.  However, when precedence information is dynamically updated, the performance of \tapnet{} is noticeably improved. 
\rh{Figure~\ref{fig:ablation_encoder}  shows some visual examples for comparison.}

\begin{table}[!t]%
	\caption{\rh{Ablation study on inputs to encoder as well as comparison to Ptr-Net~\cite{vinyals15} on RAND dataset. Our Set2Seq network can produce better performance than the Seq2Seq network Ptr-Net under the same settings. Moreover, making use of dynamic precedence information further improves \tapnet{}'s packing performance, especially on the Compactness $C$.}}
	\label{tab:comp_ptr}
	\begin{minipage}{\columnwidth}
		\begin{center}
		\begin{tabular}{l||l||l|l|l|l}
			\hline\hline
			\textbf{Input} & \textbf{Method} &  \bm{$C$} & \bm{$P$}&  \bm{$S$} &  \bm{$R$}  \\ \hline \hline 

			\multirow{2}{*}{{Shape only}} 	&  Ptr-Net &0.824 & 0.961 &  0.972 &  0.919  \\  \cline{2-6} 
			& \tapnet{}& 0.867 & 0.977 &  0.986 &  0.943  \\ \hline \hline 
			\multirow{2}{*}{{Shape + init P}} & Ptr-Net & 0.824 & 0.960 & 0.971 & 0.918 \\  \cline{2-6} 
			& \tapnet &0.859 & 0.974 &  0.986 &  0.940 \\  \hline \hline 
			\multirow{1}{*}{{Shape + dyn P}} & \textbf{\tapnet{}} & \textbf{0.925}    & \textbf{0.988} & \textbf{0.997}  & \textbf{0.970} \\ \hline \hline  

		\end{tabular}
		\end{center}
	\end{minipage}
\end{table}%

\begin{figure}[!t]
    \centering
    \includegraphics[width=0.49\textwidth]{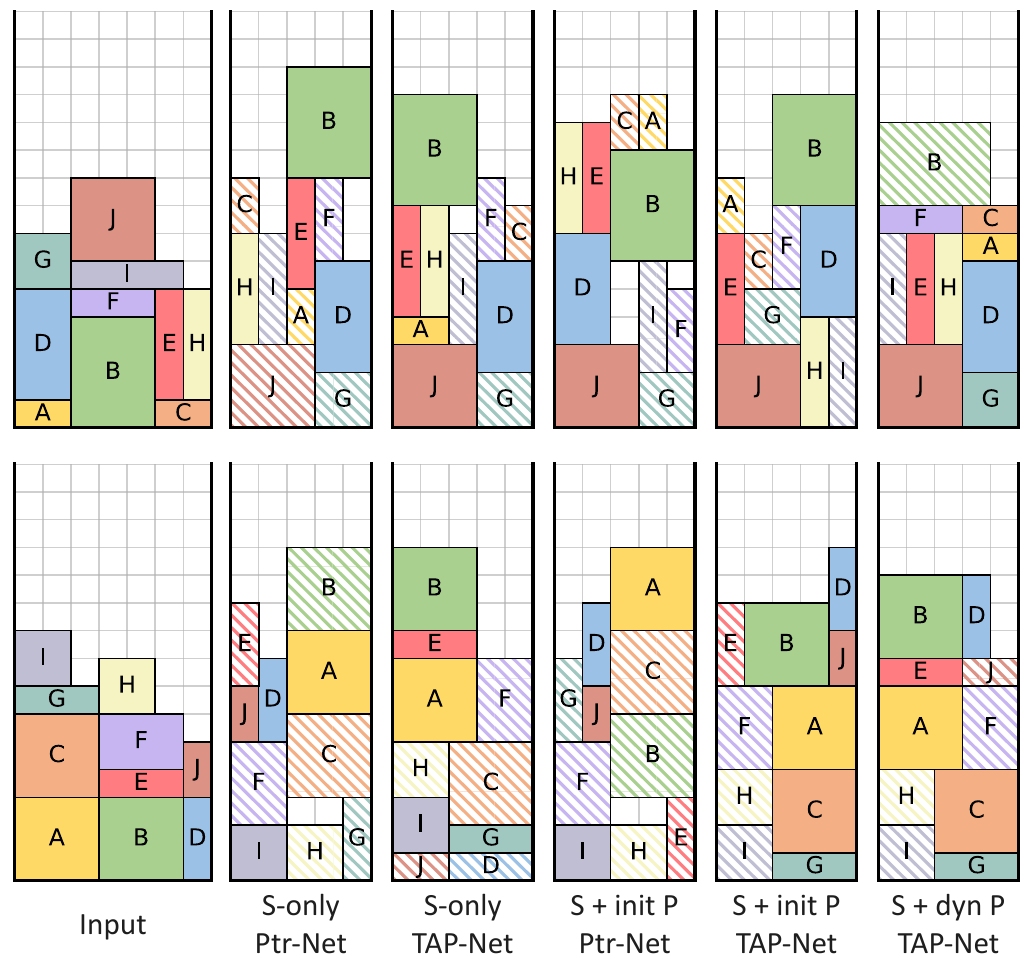}
\caption{\rh{Example results on ablation study on the inputs to encoder.}}

\label{fig:ablation_encoder}
\end{figure}

}

\paragraph{Ablation study on inputs to decoder}
A key difference between \tapnet{} architecture and the one used in~\cite{nazari18} is the input to RNN decoder. Instead of passing only the output from previous step, which is the static box shape information in our case, we further feed the information of current packing state, i.e., the current height map inside the target container.  To evaluate the necessity of both information, an ablation study is conducted using both \emph{RAND} and \emph{PPSG} datasets.

\begin{figure}[!t]
    \centering
    \includegraphics[width=0.49\textwidth]{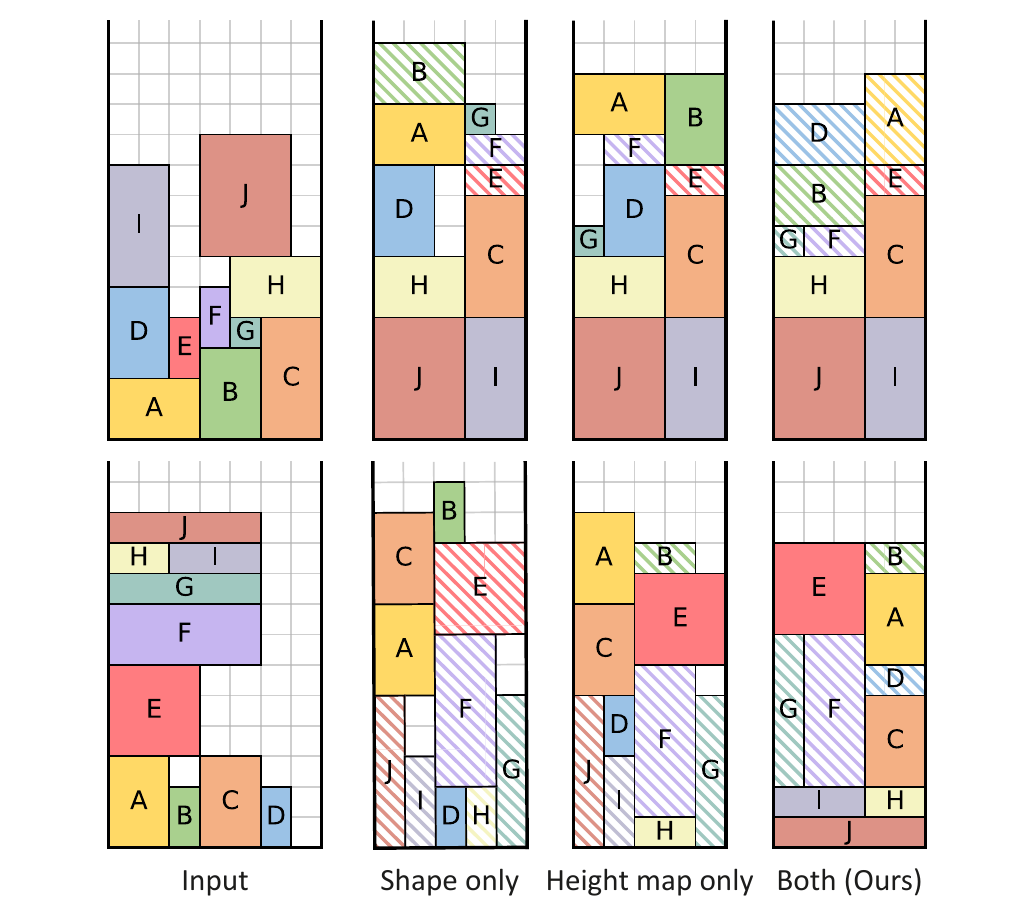}
\caption{\rh{Example results on ablation study on the inputs to decoder.}}

\label{fig:ablation_decoder}
\end{figure}

As shown in Table~\ref{tab:comp_heightmap}, on both datasets, adding height map information provides noticeable performance improvements, especially in terms of Compactness measure $C$.
Several packing examples are also shown in Figure~\ref{fig:ablation_decoder} for visual comparison. We can see that passing both shape and height map as intermediate state information to the decorder leads to better solutions, especially for samples in the \emph{PPSG} dataset.  

\begin{table}[!t]%
	\caption{Ablation study on inputs to RNN decoder on both RAND and PPSG datasets. Feeding height map information to decoder can noticeably improve packing performance, especially the Compactness $C$.} 
	\label{tab:comp_heightmap}
	\begin{minipage}{\columnwidth}
		\begin{center}
		\begin{tabular}{l||l||l|l|l|l}
			\hline\hline
			\textbf{Data} & \textbf{Decoder input} &   \bm{$C$} & \bm{$P$}&  \bm{$S$} &  \bm{$R$}    \\ \hline \hline
			
			\multirow{3}{*}{{RAND}} 
			& Shape only &0.891 & 0.983 &  0.990 &  0.954  \\ \cline{2-6} 
			& Height map only & 0.914 & 0.987 & 0.995 &  0.965 \\  \cline{2-6} 
			& \textbf{Both (Ours)} & \textbf{0.925} & \textbf{0.988} & \textbf{0.997} & \textbf{0.970}   \\  \hline \hline
			
			\multirow{3}{*}{{PPSG}} 
			& Shape only& 0.876 & 0.972 &  0.979 &  0.942   \\ \cline{2-6} 
			& Height map only&0.929 & 0.987 &  0.994 &  0.970   \\  \cline{2-6} 
			& \textbf{Both (Ours)} & \textbf{0.981} & \textbf{0.996} & \textbf{0.999} & \textbf{0.992} \\ \hline \hline
		\end{tabular}
		\end{center}
	\end{minipage}
\end{table}%
 
\begin{figure*}[!t]
    \centering
    \includegraphics[width=0.99\textwidth]{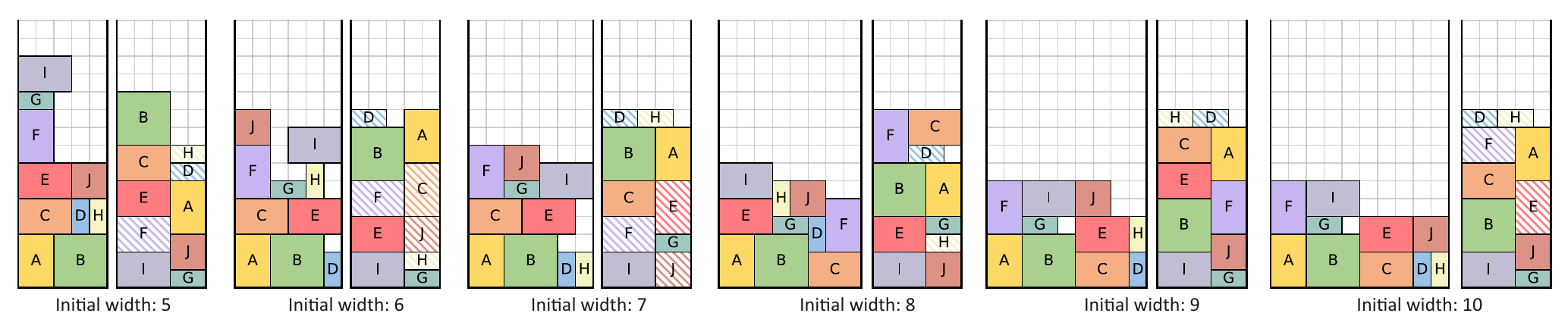}
\caption{
Packing results under different initial container widths. The packing qualities are almost identical when width is larger than or equal to 6.}
\label{fig:comp_init}
\end{figure*}

\paragraph{Impact of initial container width}
The aforementioned RAND and PPSG sets are generated by placing a set of boxes into an initial container of $W=7$. As mentioned above, the width of the container affects how likely different boxes block each other. To evaluate its impact on the final packing performance, we also generate different variants of RAND datasets by varying container width from 5 to 10. The \tapnet{} is tested on these datasets without retraining and the results are shown in Table~\ref{tab:diff_init_sizes}.

The results show that the best performance is obtained under $W=7$ setting.  Since a smaller initial container width leads to more edges in the precedence graph, the number of boxes can be selected at each step becomes limited, resulting in a noticeable performance drop. On the other hand, larger container width leads to more choices at each step.  There is slight performance drop in these cases, which is likely because \tapnet{} is not trained under these width settings.
Figure~\ref{fig:comp_init} provides visual comparison on packing results generated on the same set of boxes but under different initial container widths.

\begin{table}[!t]%
	\caption{Impact of initial container width on final packing performance. Best result is obtained under width=7, which \tapnet{} is trained for.}
	\label{tab:diff_init_sizes}
	\begin{minipage}{\columnwidth}
		\begin{center}
		\begin{tabular}{c||l|l|l|l}
			\hline\hline
			\textbf{Init container size} &  \bm{$C$} & \bm{$P$}&  \bm{$S$} &  \bm{$R$}   \\ \hline\hline
			5   & 0.872 & 0.974 & 0.984 & 0.943 \\ \hline
			6   & 0.876 & 0.974 & 0.986 & 0.945 \\ \hline
			\textbf{7}   & \textbf{0.925} & \textbf{0.988} & \textbf{0.997} &  \textbf{0.970}\\  \hline
			8   & 0.897 & 0.981 & 0.988 & 0.955 \\ \hline
			9   & 0.894 & 0.981 & 0.989 & 0.955 \\ \hline
			10 & 0.897 & 0.982 &  0.990 & 0.956 \\ \hline \hline
		\end{tabular}
		\end{center}
	\end{minipage}
\end{table}%

 \rh{
 
\begin{figure}[tb]
    \centering
    \includegraphics[width=0.99\linewidth]{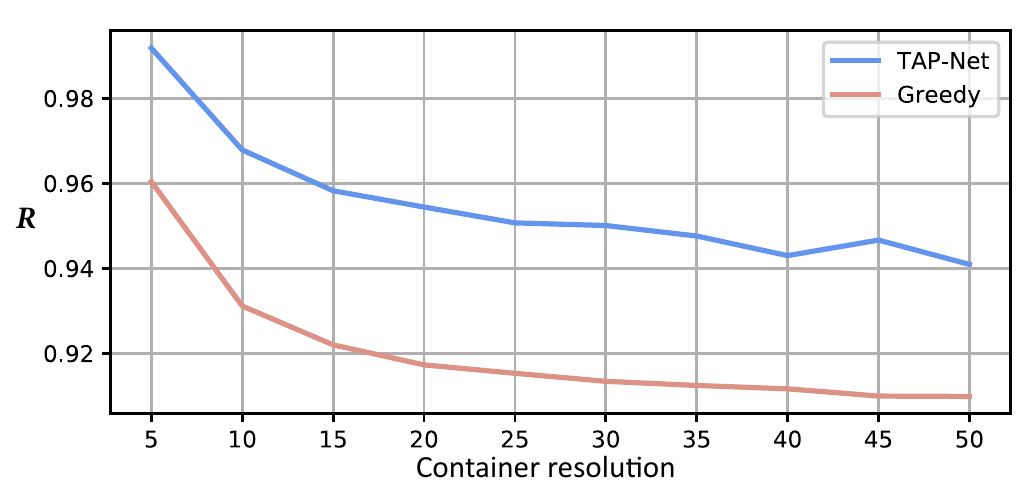}
	\caption{\change{The performance of \tapnet{} decreases slowly when the resolution increases, but is consistently better than the Greedy method.}}
	\label{fig:reso_curve}
\end{figure}

\begin{figure}[!t]
    \centering
    \includegraphics[width=0.48\textwidth]{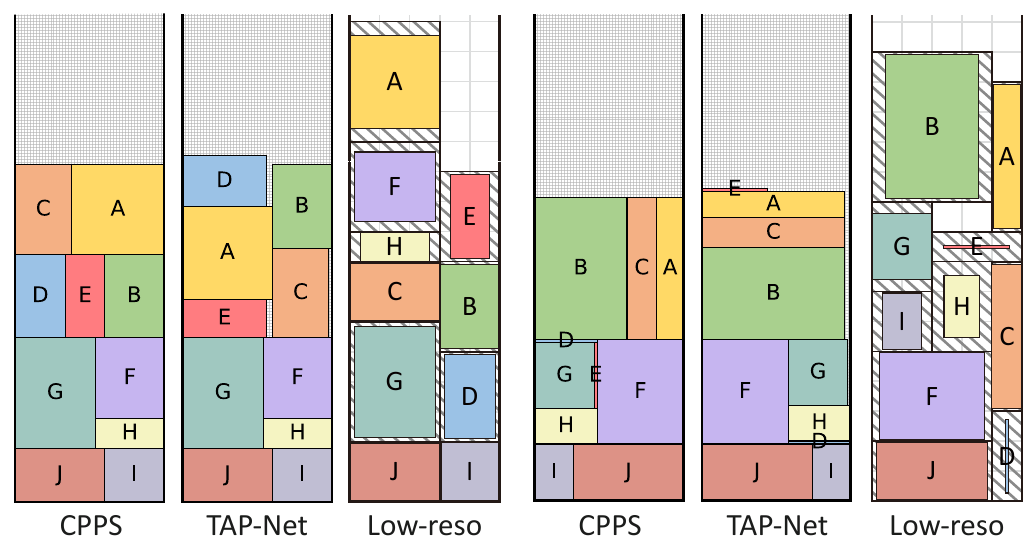}
\caption{
\rh{Packing results under high quantization resolution. The results generated by \tapnet{} are close to ideal solution (CPPS). However, when the same set of objects are quantized under low resolution, the results are less compact due to the rough approximation}.
}

\label{fig:high_res}
\end{figure}

\paragraph{Impact of quantization resolution}
For algorithm simplicity, the sizes of both container and packing objects are assumed to be integers.  When packing real world objects with arbitrary dimensions, we need to quantize their sizes using integers. Here, the resolution used for quantization can affect the performances of the packing algorithm. To evaluate this impact and how applicable \tapnet{} is in real-world applications, we trained and tested \tapnet{} with \change{quantization resolution increasing from 5 to 50, and generate boxes with corresponding sizes. For example, when the container width is set to $W=50$, the sizes of packing objects are randomly generated as integers within $[1,50]$. 
Figure~\ref{fig:reso_curve} shows that the performance of \tapnet{} will decrease when the resolution increases, but not significantly: it drops from 0.99 slowly to 0.94, when using the same amount of training data. 
Nevertheless, our method is consistently better than the greedy method. }

It is worth noting that if the number of object remains the same and only quantization resolution increases, the training and testing time won't change much.  This is because: 1) the complexity of precedence graph depends only on the number of objects, not on the quantization resolution; and 2) the processing time of the LB packing placement strategy does not depend on the number of possible placement locations which increases with  quantization resolution. In practice, the training time is around 460s per epoch, which is very similar to training for low resolution cases, and the testing time is around 3ms per sample.

Figure~\ref{fig:high_res} shows two example results generated under high quantization resolution and the corresponding results under coarse quantization.  It shows that \tapnet{} can densely pack objects under high quantization resolution and the packing density is close to the ideal CPPS cases. When the same set of objects are approximated under low quantization resolution, however, the packing results are less compact \rh{due to the rough approximation}.

}

\subsection{Results of extended applications}

\change{
\paragraph{Results on classical bin packing problem}
The classical bin packing problem can be considered as a special case of TAP, where the width of initial container is large enough so that there are no precedence constraints. Compared to previous RL-based methods for classical packing, e.g., ~\cite{hu17}, one of our key improvements is that we propose to use height maps to represent the intermediate packing states to guide the object and orientation selection, which has been shown to improve the results significantly for the TAP problem.  Here, we trained and tested our \tapnet{} on the classical packing problem and compare the performance to the Ptr-Net used in~\cite{hu17}.
As shown in Table ~\ref{tab:no_precedence}, we can see that using a Set2Seq network already outperforms Ptr-Net, and \tapnet{} that takes height maps as the extra input gets the best performance.
\begin{table}[!t]%
	\caption{\change{Packing performance obtained when applying \tapnet{}  on classical packing problem where no precedence constraints are considered, with comparison to Ptr-Ne used in~\cite{hu17}.}}
	\label{tab:no_precedence}
	\begin{minipage}{\columnwidth}
		\begin{center}
		\begin{tabular}{l||l|l|l|l}
			\hline\hline
			\textbf{Method} & \bm{$C$} & \bm{$P$}& \bm{$S$} & \bm{$R$}  \\ \hline\hline
                ~\cite{hu17}   & 0.896 & 0.988 & 0.998 & 0.961 \\ \hline
                 \textbf{\tapnet{}} & \textbf{0.948} & \textbf{0.996} & \textbf{0.999} & \textbf{0.981} \\ \hline\hline
		\end{tabular}
		\end{center}
	\end{minipage}
\end{table}%

}

\begin{table}[!t]%
	\caption{Packing performance obtained using the \tapnet{} trained on 10 objects but applied to more boxes \change{, with comparison to the Greedy method.}}
	\label{tab:larger_instances}
	\begin{minipage}{\columnwidth}
		\begin{center}
		\begin{tabular}{c||l||l|l|l|l}
			\hline\hline
			\textbf{Instance size} & \textbf{Method} & \bm{$C$} & \bm{$P$}&  \bm{$S$} &  \bm{$R$}   \\ \hline\hline
			\multirow{2}{*}{{10}}
				& \tapnet{} & \textbf{0.925} & \textbf{0.988} & \textbf{0.997} & \textbf{0.970} \\ \cline{2-6}
				& Greedy	& 0.816 & 0.972 & 0.978 & 0.922 \\ \hline \hline
			\multirow{2}{*}{{20}}
				& \tapnet{} & \textbf{0.896} & \textbf{0.962} & \textbf{0.978} & \textbf{0.945} \\ \cline{2-6}
				& Greedy	& 0.828 & 0.954 & 0.969 & 0.917 \\ \hline \hline
			\multirow{2}{*}{{30}}
				& \tapnet{} & \textbf{0.893} & \textbf{0.951} & \textbf{0.973} & \textbf{0.939} \\ \cline{2-6}
				& Greedy	& 0.828 & 0.942 & 0.965 & 0.912 \\ \hline \hline
			\multirow{2}{*}{{40}}
				& \tapnet{} & \textbf{0.927} & \textbf{0.966} & \textbf{0.980} & \textbf{0.958} \\ \cline{2-6}
				& Greedy	& 0.824 & 0.933 & 0.963 & 0.907 \\ \hline \hline
			\multirow{2}{*}{{50}}
				& \tapnet{} & \textbf{0.926} & \textbf{0.962} & \textbf{0.979} & \textbf{0.956} \\ \cline{2-6}
				& Greedy	& 0.819 & 0.925 & 0.963 & 0.902 \\ \hline \hline
		\end{tabular}
		\end{center}
	\end{minipage}
\end{table}%

\paragraph{Results on larger instance set}
By using the rolling-scheme described in Section~\ref{sec:extend_larger}, our method can be easily extended to handle larger instance sets. Using the network trained on sets with 10 objects, we tested on datasets with 20, 30, 40 and 50 objects. From the comparison shown in Table~\ref{tab:larger_instances}, we can see that with increasing the number of instances, there is slight drop in stability and pyramidality.  However, the compactness has increased, resulting in a stable overall  performance.

\begin{table}[!t]%
	\caption{Packing performance obtained when packing objects into multi-containers. Our approach is noticeably better than the baseline approach. }
	\label{tab:multi_containers}
	\begin{minipage}{\columnwidth}
		\begin{center}
		\begin{tabular}{c||l|l|l|l|l}
			\hline\hline
			\textbf{Method} &  \bm{$C$} & \bm{$P$}&  \bm{$S$} &  \bm{$R$}  &  \bm{$t$}(ms)  \\ \hline\hline
			Random & 0.732 & 0.897 & 0.947 & 0.858 &  20 \\ \hline
			Greedy & 0.745 & 0.925 & 0.954 & 0.875 & 145 \\ \hline
			\textbf{\tapnet{}} & \textbf{0.849} & \textbf{0.971} & \textbf{0.982} & \textbf{0.934} & \textbf{10} \\ \hline\hline
		\end{tabular}
		\end{center}
	\end{minipage}
\end{table}%

\begin{figure}[!t]
    \centering
    \includegraphics[width=0.46\textwidth]{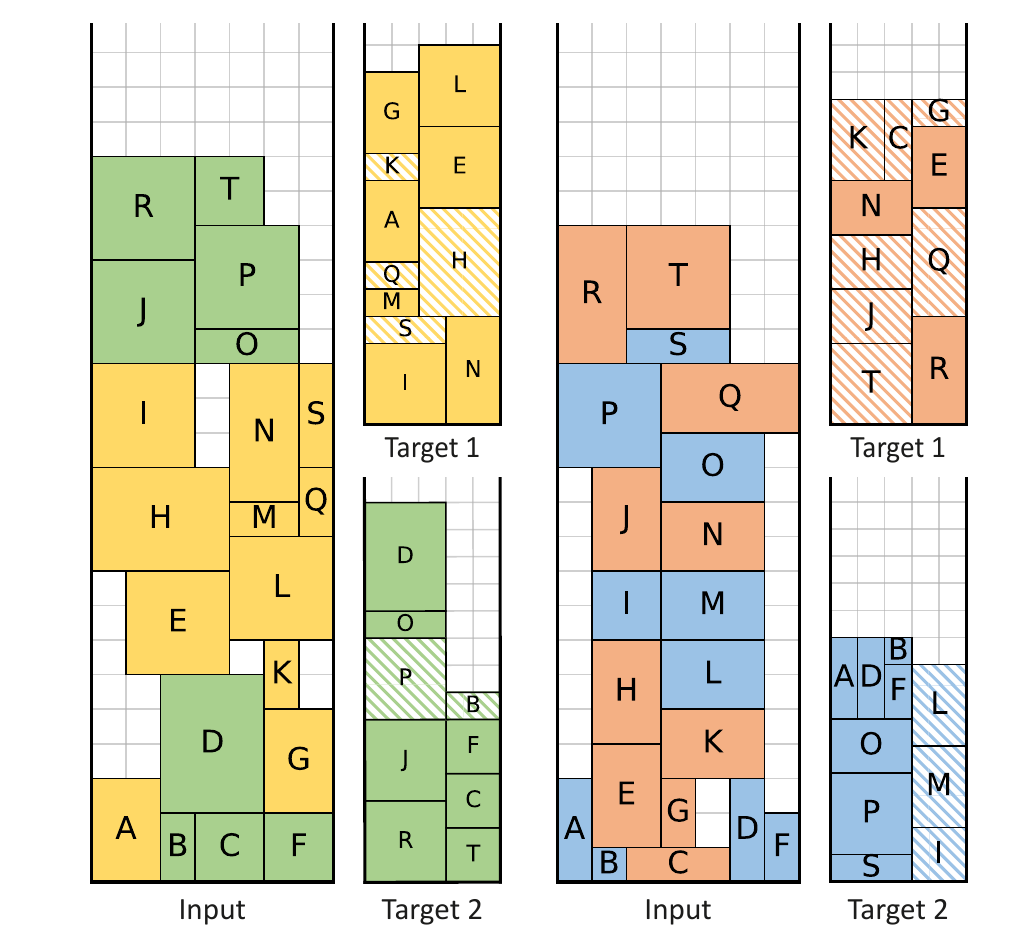}
\caption{Results obtained for packing objects into two different containers. }
\label{fig:result_multi}
\end{figure}

\paragraph{Results on multi-containers}
Our method can also be easily extended to handle cases with multiple target containers as discussed in Section~\ref{sec:extend_multi}. We test the case with two target containers on \emph{RAND}  dataset, but increase the number of input boxes to 20. The target container index for those 20 objects are randomly assigned. Comparison of the performance with baseline approach is shown in Table~\ref{tab:multi_containers}. As expected, our approach achieves much better performance than the baseline approach.  The overall measure is also close to the one achieved by packing 10 objects into a single container.


\paragraph{Results on 3D cases}
Using the same approach discussed in Subsection~\ref{sec:data}, we also generate 3D version of \emph{RAND} and \emph{PPSG} datasets. They are used to compare the performance of \tapnet{} and two baseline methods; see Table~\ref{tab:3D_packing}. Please note that \tapnet{} is retrained here since the number of object states in 3D are different and the height maps become 2D.

\begin{figure}[!t]
    \centering
    \includegraphics[width=0.46\textwidth]{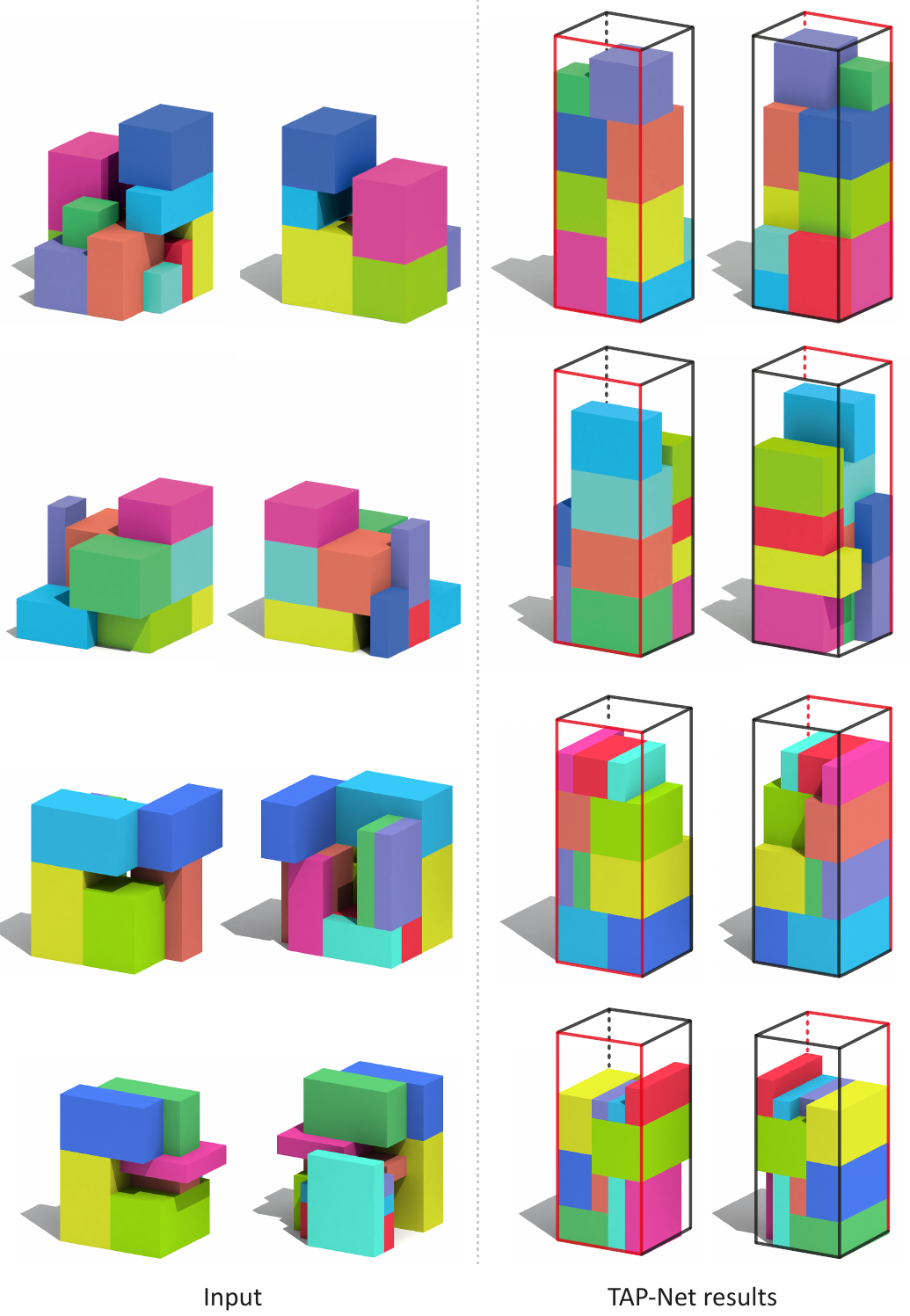}
\caption{
Example results on 3D cases: two from the RAND set (top) and two from the PPSG set (bottom).}
\label{fig:result_3D}
\end{figure}

Compared to results obtained on 2D datasets, the advantage of \tapnet{} over baseline algorithm is even more noticeable. For example, the Compactness measure between \tapnet{} and Random differs 27\% for 2D PPSG dataset and 56\% for 3D PPSG dataset.  On the other hand, due to the increased complexity with higher dimension, the average $C$ measure of results generated by \tapnet{} deviates more from the optimal solution, although the $P$ and $S$ measures do not see significant drop. Several example results obtained by our method are shown in Figure~\ref{fig:result_3D}.

\begin{table}[!t]%
	\caption{Performance comparison between \tapnet{} and the two baseline algorithms on 3D RAND and PPSG datasets.}
	\label{tab:3D_packing}
	\begin{minipage}{\columnwidth}
		\begin{center}
		\begin{tabular}{l||l||l|l|l|l|l}
			\hline\hline
			\textbf{Data} & \textbf{Method} &  \bm{$C$} & \bm{$P$}&  \bm{$S$} &  \bm{$R$}  &  \bm{$t$}(ms)  \\ \hline \hline 
			
			\multirow{3}{*}{{RAND}}
			& Random & 0.540 & 0.770 & 0.884 & 0.731 &  42 \\ \cline{2-7} 
			& Greedy & 0.656 & 0.940 & 0.978 & 0.858 & 710 \\ \cline{2-7} 
			& \textbf{\tapnet{}} & \textbf{0.764} & \textbf{0.950} & \textbf{0.996} & \textbf{0.903} & \textbf{5} \\ \hline \hline 
			
			\multirow{3}{*}{{PPSG}}
			& Random & 0.562 & 0.706 & 0.805 & 0.691 &  42 \\ \cline{2-7} 
			& Greedy & 0.870 & 0.960 & 0.990 & 0.940 & 658 \\ \cline{2-7} 
			& \textbf{\tapnet{}} & \textbf{0.877} & \textbf{0.972} & \textbf{0.997} & \textbf{0.949} & \textbf{5} \\ \hline \hline 

		\end{tabular}
		\end{center}
	\end{minipage}
\end{table}%

\section{Conclusion and future work}
\label{sec:future}

We introduce \rh{a new instance of the classical box packing problem,} transport-and-pack (TAP), and present \tapnet{}, a neural optimization model to  efficiently solve the problem. \rh{\tapnet{} is trained, via reinforcement learning, to produce an optimal sequence of box selections and orientations, starting from an initial box arrangement, and pack them into a target container. Through extensive experiments, we demonstrate that our network is more efficient than baseline schemes relying on heuristic box selections, as well as alternative network designs. Furthermore, \tapnet{} produces high-quality packing results and generalizes well to problem instances whose sizes are larger than that of the training set.}

As a first solution attempt, our current model still has several limitations. \rh{To start, we abstract the input objects by AABBs and discretize the box sizes as integers to simplify the search. While the problem remains NP-hard even with such simplifications, one negative consequence is that our packing reward can only approximate the actual packing efficiency and stability. If the objects contained in the AABBs have irregular shapes, then extra support material may need to be added between the objects to simulate packing.}

In addition, the dynamic precedence set of each object state is currently encoded using binary codes with the same size as the input object set; it is then passed to several convolutional layers for feature extraction. This information can also be considered as a sequence and encoded using an RNN, which may be more efficient and flexible when dealing with input with different sizes. 

\tapnet{} only optimizes the packing order and box orientations, while object placement into the target container is determined by a heuristic packing strategy. Althought our experiments have shown that the network appears to be able to learn to select good packing order and orientations to fit well to the packing strategy, especially when the packing strategy is simple and predictable, we still believe that it would be interesting to explore ways to solve a {\em coupled\/} problem by jointly learning or optimizing the packing order {\em and\/} the packing strategy together. \rh{More sophisticated network architecture and reward policies than our current settings may be necessary.}

\change{Besides addressing the limitations above, we are interested in extensions to transporting-and-packing objects of arbitrary shapes.
For example, we have shown that our approach scales well with the resolution of the grid, so one possible solution is that we approximate the shapes with high quantization resolution to determine the rough layout of the objects first using our approach, and then the actual object shapes are further used to guide a local refinement to make the packing more compact. A similar strategy can also be used to solve classical packing problems like texture charts packing.}
We would also like to explore the use of our \rh{neural search strategy, and its extensions, on other hard problems involving sequential 
decision making, in different application domains such as assembly-based modeling and manufacturing and robotics.}

\begin{acks}
We thank the anonymous reviewers for their valuable comments. This work was supported in part by NSFC (61872250, 61861130365, 61761146002), GD Talent Program (2019JC05X328), GD Science and Technology Program (2020A0505100064, 2018KZDXM058), LHTD (20170003), NSERC Canada (611370, 293127), gift funds from Adobe, a Google Faculty Research Award, National Engineering Laboratory for Big Data System Computing Technology, and Guangdong Laboratory of Artificial Intelligence and Digital Economy (SZ).
\end{acks}

\bibliographystyle{ACM-Reference-Format}
\bibliography{references}

\end{document}